\documentclass[%
  reprint,
 twocolumn,
superscriptaddress,
 amsmath,amssymb,
 aps,
pra,
floatfix,
notitlepage
]{revtex4-1}

\usepackage{upgreek}
\usepackage{siunitx}
\usepackage{comment}
\setcitestyle{round}

\usepackage{graphicx}
\usepackage{dcolumn}
\usepackage{bm}
\usepackage[version=3]{mhchem}
\usepackage{nicefrac}

\usepackage{color}
\usepackage{dcolumn}
\newcommand{\lio}{\ce{LiInCr4O8}}
\newcommand{\lis}{\ce{LiInCr4S8}}
\newcommand{\lgo}{\ce{LiGaCr4O8}}
\newcommand{\lgs}{\ce{LiGaCr4S8}}
\newcommand{\cis}{\ce{CuInCr4S8}}
\newcommand{\cise}{\ce{CuInCr4Se8}}

\usepackage{pdfpages}
\makeatletter
\AtBeginDocument{\let\LS@rot\@undefined}
\makeatother

\begin{document}


\title{Breathing chromium spinels: a showcase for a variety of pyrochlore Heisenberg Hamiltonians
}

\author{Pratyay Ghosh}
\affiliation{Department of Physics, Indian Institute of Technology Madras, Chennai 600036, India}

\author{Yasir Iqbal}
\affiliation{Department of Physics, Indian Institute of Technology Madras, Chennai 600036, India}

\author{Tobias M\"uller}
\affiliation{Institute for Theoretical Physics and Astrophysics, Julius-Maximilians-Universit\"at W\"urzburg, Am Hubland, 97074 W\"urzburg, Germany}

\author{Ravi T. Ponnaganti}
\affiliation{Department of Physics, Indian Institute of Technology Madras, Chennai 600036, India}

\author{Ronny Thomale}
\affiliation{Institute for Theoretical Physics and Astrophysics, Julius-Maximilians-Universit\"at W\"urzburg, Am Hubland, 97074 W\"urzburg, Germany}

\author{Rajesh Narayanan}
\affiliation{Department of Physics, Indian Institute of Technology Madras, Chennai 600036, India}

\author{Johannes Reuther}
\affiliation{Dahlem Center for Complex Quantum Systems and Fachbereich Physik, Freie Universit\"at Berlin, 14195 Berlin, Germany}

\author{Michel J. P. Gingras}
\affiliation{Department of Physics and Astronomy, University of Waterloo, Waterloo, Ontario, N2L 3G1, Canada}
\affiliation{Quantum Materials Program, Canadian Institute for Advanced Research, MaRS Centre,
West Tower 661 University Ave., Suite 505, Toronto, ON, M5G 1M1, Canada}

\author{Harald O. Jeschke$^*$}
\affiliation{Research Institute for Interdisciplinary Science, Okayama University, Okayama 700-8530, Japan}
\email{jeschke@okayama-u.ac.jp}

\date{\today}

\begin{abstract}
We address the long-standing problem of the microscopic origin of the richly diverse phenomena in the chromium breathing pyrochlore material family. Combining electronic structure and renormalization group techniques we resolve the magnetic interactions and analyze their reciprocal-space susceptibility. We show that the physics of these materials is principally governed by long-range Heisenberg Hamiltonian interactions, a hitherto unappreciated fact. Our calculations uncover that in these isostructural compounds, the choice of chalcogen triggers a proximity of the materials to classical spin liquids featuring degenerate manifolds of wave-vectors of different dimensions: A Coulomb phase with three-dimensional degeneracy for {\lio} and {\lgo}, a spiral spin liquid with two-dimensional degeneracy for {\cise} and one-dimensional line degeneracies characteristic of the face-centered cubic antiferromagnet for {\lis}, {\lgs} and {\cis}. The surprisingly complex array of prototypical pyrochlore behaviors we discovered in chromium spinels may inspire studies of transition paths between different semi-classical spin liquids by doping or pressure.
\end{abstract}


\maketitle

\noindent{\bf Introduction}

\noindent Over the past thirty years, materials with magnetic moments on the vertices of networks of corner-shared triangular or tetrahedral units have played center stage in the experimental search for exotic magnetic states driven by  competing, i.e. frustrated, interactions~\cite{Lacroix2011}. In this context, the pyrochlore lattice of corner-shared tetrahedra has emerged as a quintessential example of high frustration in three dimensions with a multitude of new phenomena having been uncovered~\cite{Gardner2010}. 

In particular, for rare-earth pyrochlore oxides with a trivalent $4f$ magnetic ion, there is a wealth of stoichiometric compounds available in large single-crystal form, necessary for detailed neutron scattering studies. This, along with the synergetic dialogue between theory and experiments, has led to the experimental discovery and the theoretical rationalization of numerous phenomena like spin ice physics, magnetic moment fragmentation, order-by-disorder, fragile splayed-ferromagnetism and candidate spin liquid phenomenology~\cite{Hallas2018,Rau2019}. 

In contrast with the above rare-earth systems characterized by strongly anisotropic interactions at the $O(1)$ K scale, there is a paucity of magnetic compounds involving transition metal ions on a pyrochlore network and coupled via a  primarily isotropic Heisenberg exchange of a high ($O(10^2)$ K)  energy  scale. The availability of such compounds in single-crystal form would empower researchers to explore and possibly discover novel collective behaviors existing over a wider and more experimentally accessible temperature window. Specifically, this would allow exposing new ways in which perturbations, which might have appeared inconsequential from a cursory assessment, ultimately emerge as the dictating forces behind the low-temperature and low-energy scale physics of the material. In that regard, the recent successful synthesis of magnetic pyrochlore fluorides $ABM_2$F$_7$ ($A=$Na; $B=$Ca, Sr; $M=$ Ni, Co, Fe, Mn) in large single crystals is indeed an exciting and most welcome development in the field of highly-frustrated magnetism~\cite{Plumb2019}. There is, however, a short-coming with the latter compounds whose implications remain to be fully ascertained: The  cation disorder on the non-magnetic $A$ and $B$ sites is expected to randomize the $M$-$M$ superexchange and, most likely, ultimately drive the low-temperature state of these systems to a semi-classical spin glass phase, albeit with some nontrivial spin dynamics~\cite{Plumb2019}. In this context, the disorder-free so-called breathing chromium spinels,  which define a fairly broad range of materials (e.g. {\lio}, {\lgo}, {\lis}, {\lgs}, {\cis} and {\cise})~\cite{Okamoto2013,Duda2008,Tanaka2014,Nilsen2015,Okamoto2015,Lee2016,Saha2016,Okamoto2017,Wawrzynczak2017,Pokharel2018,Okamoto2018}, constitute a significant opportunity for carrying out the  above research program in and below the $O(10^2)$ K temperature scale.

\begin{figure}[t]
\includegraphics[width=0.49\textwidth]{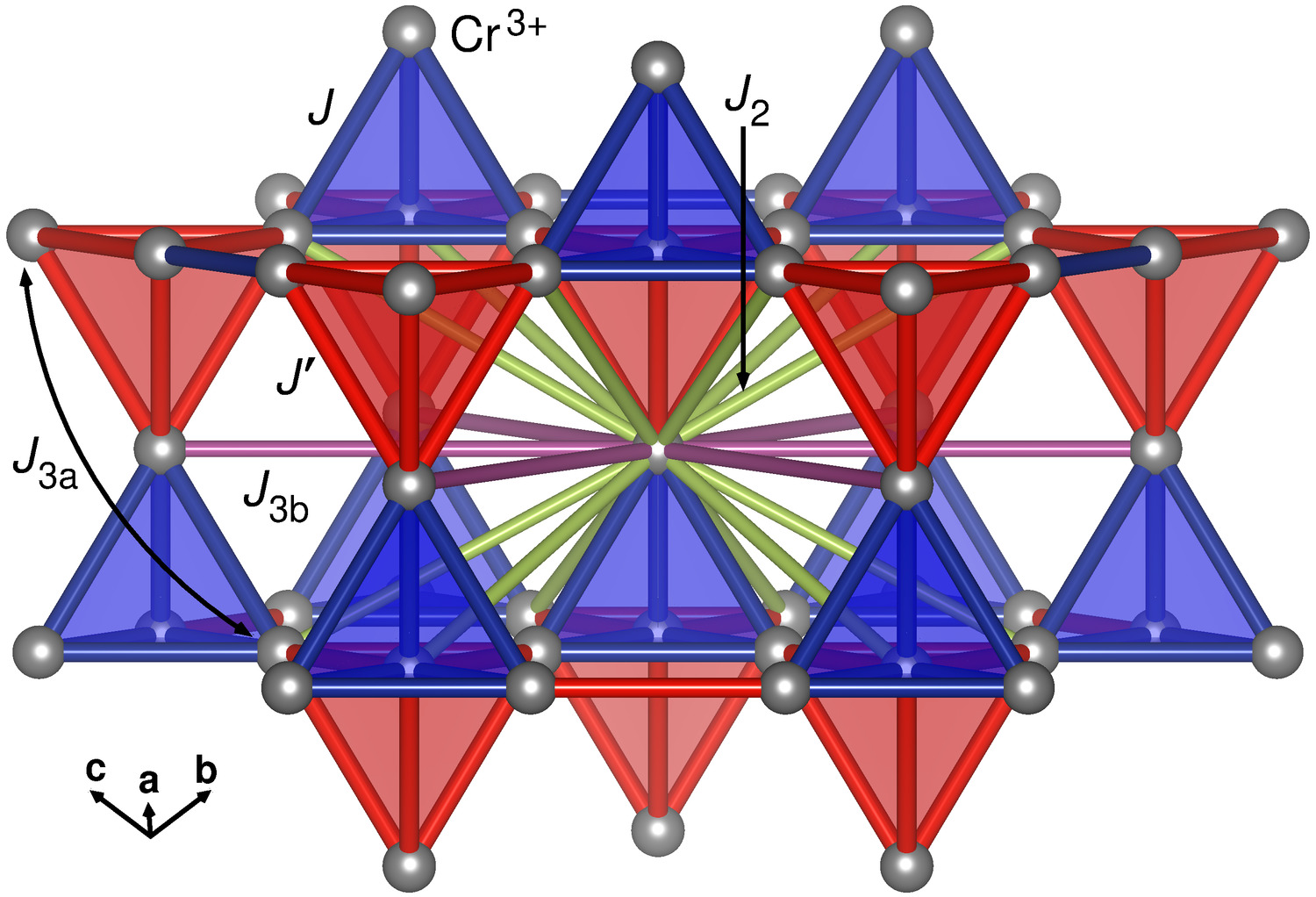}
\caption{{\bf Section of the breathing pyrochlore structure.} First, second and third nearest neighbors are highlighted for the central Cr$^{3+}$ ion. We use $J$, $J'$ to label the two nearest-neighbor exchange paths within small (blue) and large (red) tetrahedra, respectively. Note that $J_2$ connects Cr$^{3+}$ ions of small and large tetrahedra so that there is only one kind of such a coupling. $J_{3a}$ and $J_{3b}$ are the two symmetry inequivalent third nearest neighbors; the $J_{3a}$ features an in-between Cr$^{3+}$ ion while the $J_{3b}$ does not.}
\label{fig:paths}
\end{figure}

The introduction of a breathing degree of freedom to the regular pyrochlore lattice, characterized by the alternation of small and large tetrahedra, results in two inequivalent nearest-neighbor exchange couplings, with $J$ for the small and $J'$ for the large tetrahedra, and adds a new level of complexity and thus richness to the regular pyrochlore Heisenberg Hamiltonian~\cite{Lacroix2011}. In the breathing chromium spinels, of chemical formula $AA'$Cr$_4$X$_8$ ($A$=Li, Cu;  $A'$=Ga, In; $X$=O, S, Se), the magnetic Cr$^{3+}$ spin $S=3/2$ resides on a breathing pyrochlore lattice (see Figure~\ref{fig:paths}). Of particular interest, thanks to their significantly different size, the ions at the $A$ and $A'$ sites do not chemically admix (i.e. they order), and the breathing chromium spinels are not subject to the $A/B$ site disorder and  consequential exchange randomness of the aforementioned $ABM_2$F$_7$ pyrochlores.   While an ordered arrangement of the  $A$ and $A'$ cations  was first reported over half-a-century ago~\cite{Joubert1966}, it is only recently that breathing chromium spinels have become a subject of focused attention~\cite{Okamoto2013}.   A variety of experimental results have been reported, including a complex sequence of structural and magnetic transitions in {\lio} and {\lgo}~\cite{Nilsen2015,Lee2016,Tanaka2014,Saha2016} as well as a half-magnetization plateau in  high magnetic field measurements on {\lio}~\cite{Okamoto2017}.

On the theoretical front, the properties of the classical nearest-neighbor breathing pyrochlore Hamiltonian have been investigated~\cite{Benton2015}, and spin-spin correlations in the ground state~\cite{Tsunetsugu2017} and the effects of spin-lattice coupling~\cite{Aoyama2019} have been explored. Possible implications of the topological aspects of magnetic excitations in the breathing pyrochlore lattice have been pointed out~\cite{Li2016,Ezawa2018}. However, to the best of our knowledge, there appears to have been no attempt to (i) provide a microscopic determination of the Cr-Cr exchange interactions and to (ii)  investigate using an unbiased theoretical framework the implications of the \emph{concurrent} thermal and quantum fluctuations operating in these compounds. Such an endeavor is necessary to gain a deeper microscopic understanding  of the already available experimental results beyond the simplest interpretation currently provided  as well as identifying    new and exciting avenues of investigation in these materials -- this is the purpose of the present work. As we shall discuss later on,  different signs (ferromagnetic or antiferromagnetic) exchange couplings $J$ and $J'$ are realized in the oxides ($X=$O), sulfide ($X=$S) and selenides ($X=$Se) breathing chromium spinels. This provide an uncharted territory for the investigation of the nontrivial role played by  long-range interactions beyond $J$ and $J'$ in these materials. This leads us to propose that breathing chromium  spinels constitute a promising platform to explore the role of strong competing long-range interactions in a semi-classical ($S=3/2$) regime.  In particular, we predict these to engender new forms of nontrivial correlations in the cooperative paramagnetic regime characterized by different  classically degenerate ground state manifolds of different dimension and intricate magnetic orders at low temperatures. Examples include emergent effective Heisenberg antiferromagnetism on the face-centered cubic lattice~\cite{Benton2015}, a Coulomb phase~\cite{Moessner1998} and a spiral spin liquid~\cite{Bergman2007}.
\vspace{0.3cm}

\begin{figure*}[htb]
\includegraphics[width=0.95\textwidth]{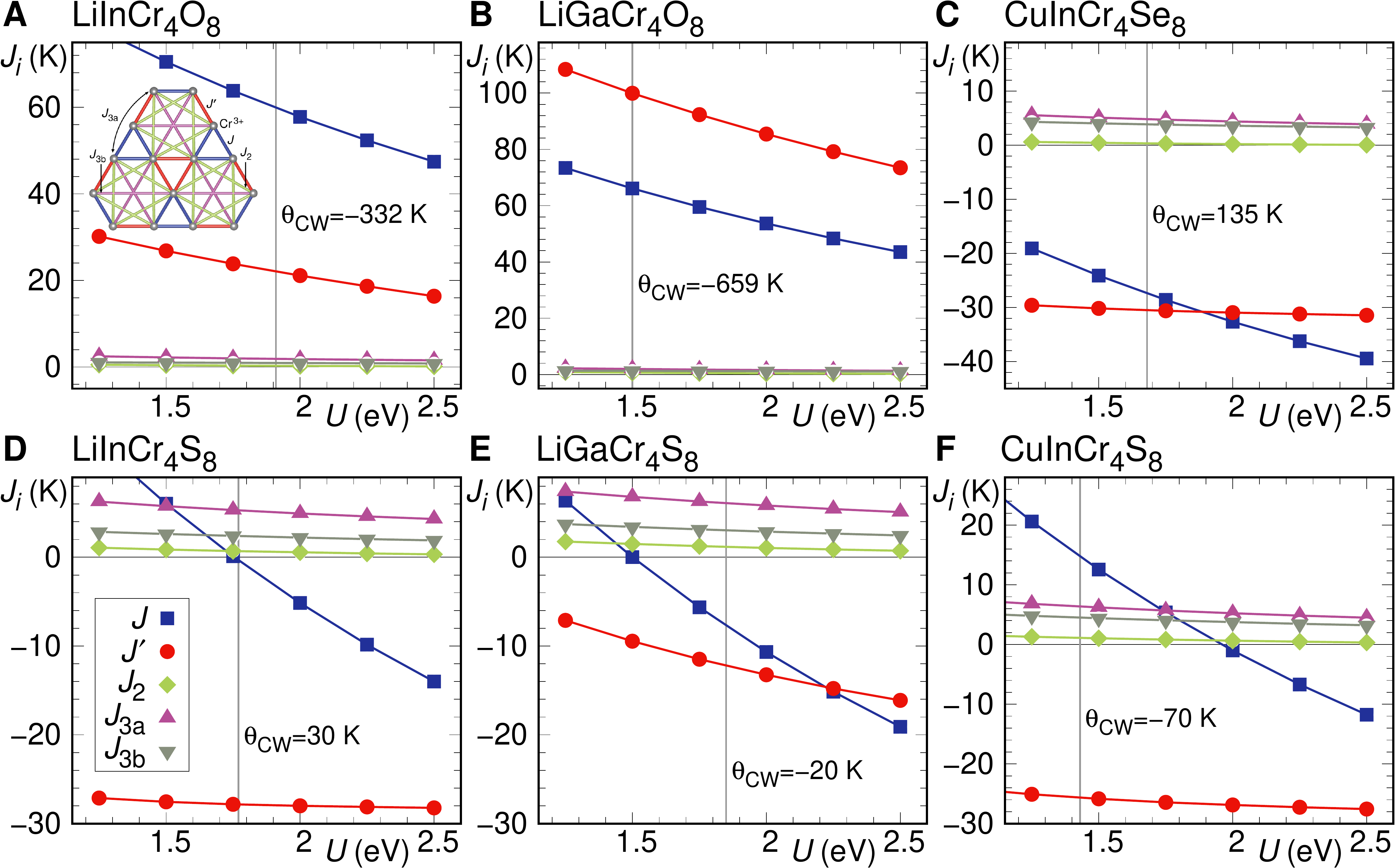}
\caption{{\bf Exchange couplings of six breathing pyrochlore compounds.} The values are calculated by mapping of DFT total energies for ({\bf A}) {\lio}, ({\bf B}) {\lgo}, ({\bf C}) {\cise}, ({\bf D}) {\lis}, ({\bf E}) {\lgs} and ({\bf F}) {\cis} and shown as function of interaction strength $U$.  The vertical lines indicate the $U$ value at which the known experimental Curie-Weiss temperature is reproduced. The inset of (A) shows a single kagome plane cut out of the pyrochlore lattice to visualize the exchange connectivity.}
\label{fig:couplings}
\end{figure*}

\begin{table*}[htb]
\caption{{\bf Exchange couplings of six chromium breathing pyrochlore compounds}. The first six lines are calculated for  for the room temperature structures, while the last two lines are for low temperature structures. All Hamiltonian parameters are calculated within GGA+$U$ with $J_{\rm H}=0.72$~eV. The values of the $J_i$ are given with respect to classical spins of length $S=3/2$ and without double counting of bonds [see Eq.~\eqref{eq:H}]. }\label{tab:couplings}
\begin{ruledtabular}
\begin{tabular}{lllllllll}
Material& T(K) & $U$~(eV)& $J$~(K)&    $J'$~(K)&  $J_{2}$~(K)&   $J_{3a}$~(K)&  $J_{3b}$~(K)& $\uptheta_{\rm CW}$~(K)\\\hline
{\lio} & RT & 1.91 & 59.8(2) &  22.0(2) &  0.3(1) &  1.9(1) &  0.9(1) & -332 \\ 
{\lgo} & RT & 1.50 & 66.2(3) &  100.0(2) &  0.7(1) &  2.0(1) &  1.3(1) & -658 \\ 
{\lis} & RT & 1.77 & -0.3(1) &  -28.0(1) &  0.7(1) &  5.3(1) &  2.4(1) & 30 \\ 
{\lgs} & RT & 1.85 & -7.7(1) &  -12.2(1) &  1.2(1) &  6.1(1) &  3.0(1) & -20 \\ 
{\cis} & RT & 1.43 & 14.7(1) &  -26.0(1) &  1.1(1) &  6.4(1) &  4.5(1) & -70 \\ 
{\cise} & RT & 1.68 & -25.4(2) &  -31.0(1) &  0.3(1) &  4.8(1) &  3.9(1) & 135 \\ \hline
{\lio} & 20 & 1.77 & 40.6(2) &  40.8(2) &  0.3(1) &  2.0(1) &  0.9(1) & -332 \\ 
{\lgs} & 10 & 1.80 &-10.8(1) &  -9.5(1) &  1.2(1) &  6.2(1) &  3.1(1) & 30 
\end{tabular}\end{ruledtabular}
\end{table*}

\begin{figure*}[htb]
\includegraphics[width=0.9\textwidth]{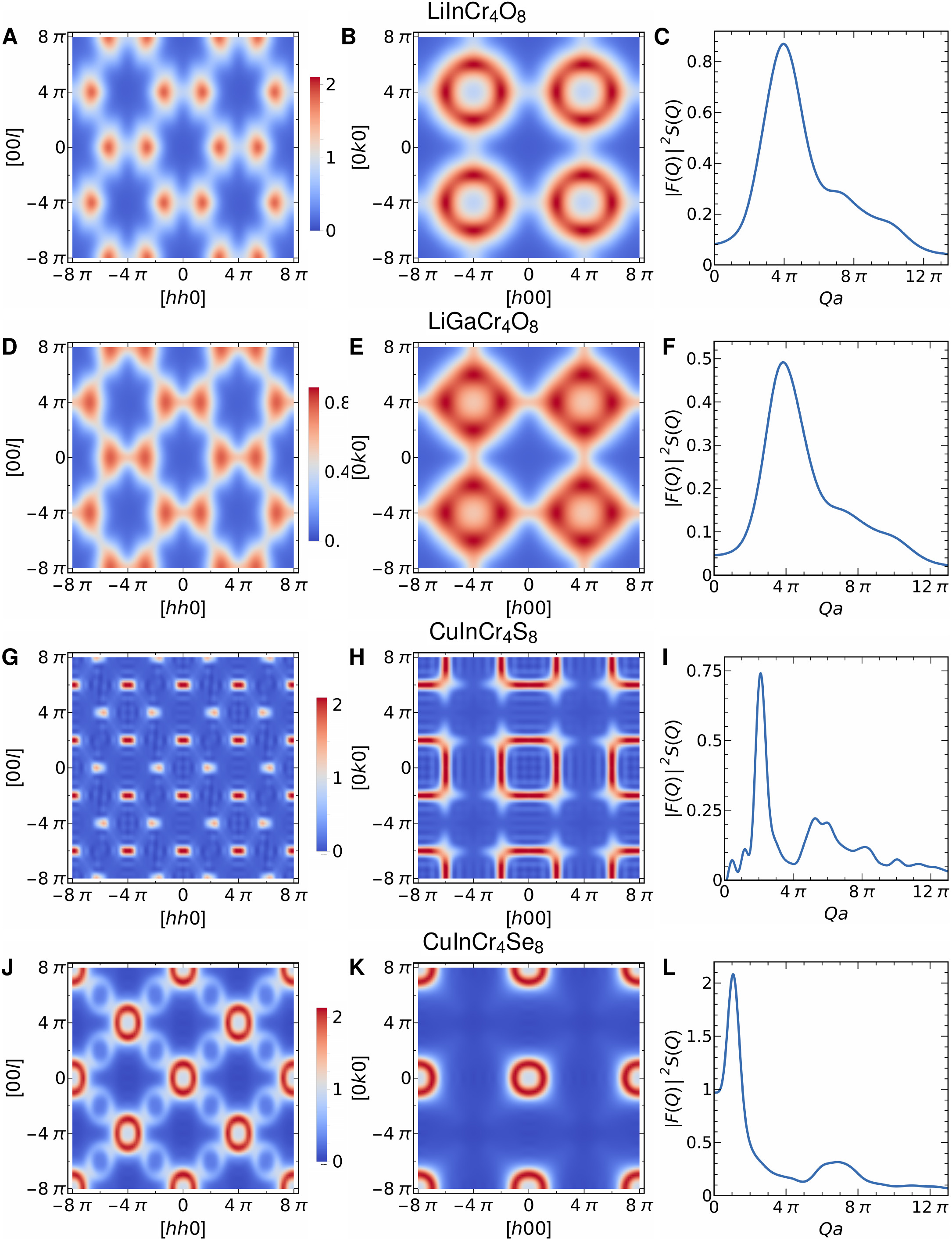}
\caption{{\bf Spin susceptibility profile for {\lio}, {\lgo}, {\cis} and {\cise} obtained using PFFRG.} This is evaluated at the breakdown point in the RG flow [marked by an arrow in SI Fig.~S2]. ({\bf A}, {\bf D}, {\bf G}, {\bf J}) Spin susceptibility projected on the $[hhl]$ plane. ({\bf B}, {\bf E}, {\bf H}, {\bf K}) Spin susceptibility projected on the $[hk0]$ plane. Susceptibility units are $1/\bar{J}$ ($\bar{J}=\sqrt{J^2+J'^2}$), and the axes are in units of inverse lattice parameter $1/a$.  ({\bf C}, {\bf F}, {\bf I}, {\bf L}) Plots of the form factor modulated powder averaged susceptibility $|F(Q)|^{2}S(Q)$ vs. $Qa$. 
}\label{fig:susceptibilityprofiles}
\end{figure*}

\begin{figure*}[htb]
\includegraphics[width=0.95\textwidth]{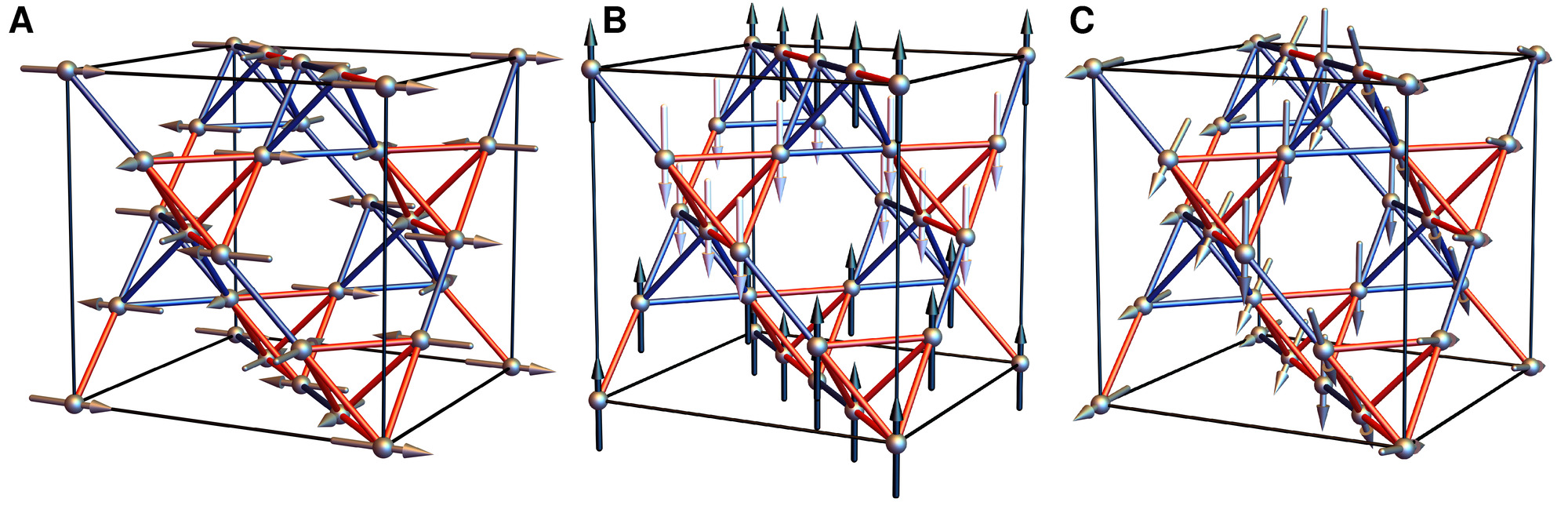}
\caption{{\bf Ground state spin configurations obtained by iterative minimization.} A conventional unit cell of the pyrochlore lattice is shown. ({\bf A}) ${\bf q}=\frac{2\pi}{a}(2,1,0)$ order relevant for the oxides {\lio} and {\lgo}. ({\bf B}) ${\bf q}=\frac{2\pi}{a}(1,0,0)$ order relevant for the sulfides {\lis}, {\lgs} and {\cis}. ({\bf C}) $(q,0,0)$ type planar incommensurate spiral relevant for {\cise}.}\label{fig:orders}
\end{figure*}

\begin{figure*}[htb]
\includegraphics[width=0.9\textwidth]{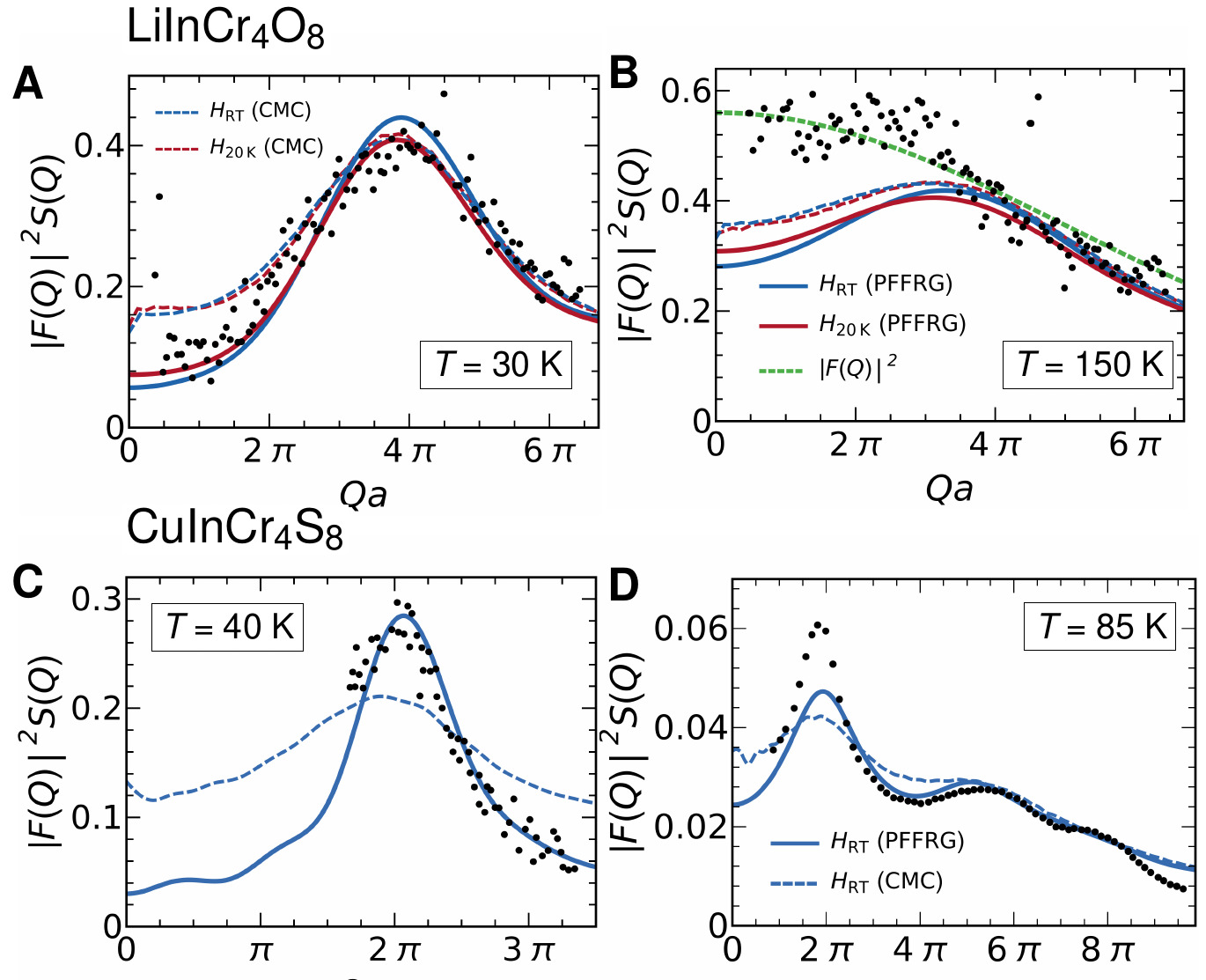}
\caption{{\bf Comparison between neutron scattering data and calculated structure factors.}
  ({\bf A},{\bf B}) Calculated structure factors $S(Q)$ from PFFRG and classical Monte Carlo for {\lio} multiplied with the form factor $|F(Q)|^2$ for Cr$^{3+}$ ions, compared to experimental neutron scattering data for a powder sample, taken from Ref.~\protect\onlinecite{Okamoto2015}. $H_{\rm RT}$ and $H_{20{\rm K}}$ refer to the Hamiltonian parameters calculated for room temperature [see Figure~\ref{fig:couplings}A] and $T=20$~K [see SI Fig.~S1A] structures, respectively. ({\bf C},{\bf D})  Plot of $|F(Q)|^{2}S(Q)$ vs $Qa$ at 40~K and 85~K, respectively, for {\cis} (blue curve). The black dots shows the experimental data obtained by Plumier {\it et al}.~\cite{Plumier1971,Plumier1977}.}\label{fig:powderaverage}
\end{figure*}

\noindent{\bf Results}

\noindent{\bf Breathing pyrochlore Hamiltonians} 

{\it Room temperature structures.-} We first perform electronic structure calculations for all presently known room temperature structures of chromium breathing spinels. We use the structures of Okamoto {\it et al.} (Ref.~\onlinecite{Okamoto2013}) for {\lio} and {\lgo}, the structures of  Okamoto {\it et al.} (Ref.~\onlinecite{Okamoto2018}) for {\lis}, {\lgs} and {\cis} and that of Duda {\it et al.} (Ref.~\onlinecite{Duda2008}) for {\cise}. The main exchange connectivity of all structures is illustrated in Figure~\ref{fig:paths}. We follow the experimental literature~\cite{Okamoto2013,Tanaka2014,Nilsen2015,Lee2016} in referring to the exchange couplings within the small and large tetrahedra as $J$ and $J'$, respectively. As in the isotropic pyrochlore lattice~\cite{Iqbal2017}, there are twelve second nearest neighbors $J_2$ and twelve third nearest neighbors which split into two symmetry inequivalent classes; six of these, named $J_{3a}$, have an in-between Cr$^{3+}$ ion, and the six others, named $J_{3b}$, do not.

Figure~\ref{fig:couplings} presents the first main result of our work. The exchange couplings as defined in the Heisenberg Hamiltonian of the form 
\begin{equation}
    H=\sum_{i<j} J_{ij}~{\bf S}_i\cdot {\bf S}_j\,.
\label{eq:H}
\end{equation}
are shown as a function of interaction strength $U$ in the GGA+$U$ functional. For all six compounds, a unique $U$ value can be determined at which the couplings yield the experimental Curie-Weiss temperature (vertical lines). These fixed parameter values lead to the sets of Hamiltonian couplings given in Table~\ref{tab:couplings}. The $U$ values are very reasonable~\cite{Tapp2017} for Cr$^{3+}$ and are all located in a narrow interval $1.43\,\text{eV}\le U\le 1.91\,\text{eV}$, indicating that the great diversity among the chromium breathing pyrochlores, varying from antiferromagnetic to ferromagnetic Heisenberg models, can all be described with roughly the same exchange correlation functional. We find that both oxides, {\lio} and {\lgo}, are dominated by antiferromagnetic $J$ and $J'$ couplings (Figures~\ref{fig:couplings}A and B), with longer range couplings being negligibly small. Interestingly, it is found that in {\lgo} the large tetrahedron is associated with the larger exchange coupling. The calculated breathing exchange anisotropy for {\lgo} is $B_f\equiv J/J' =0.66$, very close to the experimental estimate $B_f=0.6$~\cite{Okamoto2013}. However, according to our calculation, {\lio} at $B_f\equiv J'/J=0.37$ is far less anisotropic than assumed so far ($B_f=0.1$)~\cite{Okamoto2013}. For the sulfides {\lis}, {\lgs} and {\cis} (Figures~\ref{fig:couplings}D-F), the large tetrahedra is characterized by ferromagnetic exchange $J'$. However, the biggest surprise here is the presence of substantial second- and third-neighbor couplings. As each Cr$^{3+}$ ion has a total of 24 such bonds, compared to six $J$ or $J'$ bonds, they can have a substantial influence on the behavior of the materials. Interestingly, the fact that the experimentally determined Curie-Weiss temperature of {\lgs} is small is thus readily explained by a cancellation of ferromagnetic $J$ and $J'$ by antiferromagnetic $J_{2}$ and $J_{3a}$, $J_{3b}$ couplings rather than by opposite signs of $J$ and $J'$ as hypothesized in Ref.~\onlinecite{Okamoto2018}. Opposite signs of $J$ and $J'$ are only found for {\cis}. Finally, the selenide {\cise} (Figure~\ref{fig:couplings}C) is dominated by ferromagnetic $J$ and $J'$ couplings; however, even for this compound, the third neighbor couplings $J_{3a}$ and $J_{3b}$ are antiferromagnetic and, most importantly, non-negligible.

{\it Temperature dependence.-} For most materials, only a room temperature structure is on record. However, Ref.~\onlinecite{Nilsen2015} gives a $T=20$~K structure for {\lio}, while Pokharel {\it et al.}~\cite{Pokharel2018} provide a $T=10$~K structure for {\lgs}. The two Cr-Cr distances within small and large tetrahedra for these two structures are tabulated along with the measurements of the six room temperature structures in SI Table~S2. While the breathing anisotropy of the oxide {\lio} decreases significantly at low temperature, it is essentially temperature independent for the sulfide {\lgs}. Studying the two cubic low temperature structures can give us an indication of how temperature dependent the electronic behavior of the breathing pyrochlore materials is within the cubic $F\bar{4}3m$ structure. Clearly, a systematic study of the temperature dependence of the exchange couplings would require more detailed structural information for the other materials. The last two lines of Table~\ref{tab:couplings} show the exchange interactions calculated for the two low temperature structures. Even though the low temperature structure of {\lio} has a reduced breathing anisotropy of $r'/r=1.029$, compared to $r'/r=1.051$ at room temperature, the exchange couplings $J$ and $J'$ become nearly identical (see SI Fig.~S1a), with the room temperature exchange anisotropy of $J'/J= 0.37$ having been eliminated. For {\lgs}, while the low temperature structural anisotropy is still significant at $r'/r=1.070$, the two ferromagnetic couplings $J$ and $J'$ become very similar near the $U$ value determined by the experimental $\uptheta_{\rm CW}$ (see SI Fig.~S1B). Thus, we find that the nearest-neighbor exchanges $J$ and $J'$ remain antiferromagnetic for {\lio} and ferromagnetic for {\lgs}, but in both cases they become very similar at low temperatures. At the same time, the further neighbor couplings remain very small for the oxide and substantial for the sulfide.

Energy mapping clearly identifies three different types of breathing pyrochlore Hamiltonians: Mostly antiferromagnetic oxides, sulfides with ferromagnetic large tetrahedron and antiferromagnetic longer range $(J_{2},J_{3\rm a},J_{3\rm b})$ exchange, and the mostly ferromagnetic selenide perturbed by crucial antiferromagnetic longer range interactions. Thus, we organize our pseudofermion functional renormalization group analysis of the Hamiltonians into oxide, sulfide and selenide sections.
\vspace{0.3cm}

\noindent{\bf Pseudofermion functional renormalization group calculations} 

\noindent{\bf Oxides}

For {\lio} and {\lgo}, the $J$ and $J'$ couplings forming the tetrahedra are \emph{both} antiferromagnetic. With only these two couplings, one would expect the typical bow-tie features with the associated pinch points in the reciprocal space susceptibility at sufficiently low temperatures (see Ref.~\onlinecite{Iqbal2019}). In this previous study, we showed that the inclusion of even very weak second nearest-neighbor couplings can dramatically alter the susceptibility profile of the nearest-neighbor ground state phase and destroy the pinch points and bow-ties~\cite{Conlon2010}. As stated above, for both breathing pyrochlore oxides, our DFT analysis reveals the presence of additional antiferromagnetic $J_2$,  $J_{3a}$, and $J_{3b}$ couplings (see Table~\ref{tab:couplings}). At the classical level, it is known that when the dominant interactions $J$ and $J'$ are large and antiferromagnetic such that the four spins at the vertices of each tetrahedra can be assumed to sum to zero approximately, then a third neighbor coupling $J_{3a}$ in an ideal pyrochlore lattice is equivalent at temperature $T\ll {\rm min}(J/k_{\rm B},J'/k_{\rm B})$ to a $J_2$ coupling of opposite sign~\cite{Chern2008,Conlon2010}. Thus, effectively, both chromium oxides are approximately described by a pyrochlore Hamiltonian with antiferromagnetic $J$, $J'$ and ferromagnetic $J_2$: $(J_2-J_{3a})/\bar{J}=-0.025$ for {\lio} and $-0.011$ for {\lgo} ($\bar{J}=\sqrt{J^2+J'^2}$ is the overall nearest-neighbor energy scale, listed in SI Table~S1. According to Ref.~\onlinecite{Iqbal2019}, we expect that the spectral weight moves away from the pinch point and forms two symmetrical maxima, resulting in a hexagonal cluster pattern of intensities in the $[hhl]$ plane. Indeed, using the exchange couplings obtained for the room temperature structure (see Table~\ref{tab:couplings}), we employ PFFRG to calculate the magnetic susceptibility profiles which are shown in  Figures~\ref{fig:susceptibilityprofiles}A-F. They are found to closely resemble the susceptibility of the isotropic $J_1$-$J_2$ pyrochlore magnet for antiferromagnetic $J_1$ and ferromagnetic $J_2$~\cite{Iqbal2019}. It is worth noting that, due to the relatively weaker effective second nearest-neighbor coupling in {\lgo}, the pinch points and bow-ties are better preserved at low temperature for {\lgo} (see Figure~\ref{fig:susceptibilityprofiles}D) in comparison to {\lio} (Figure~\ref{fig:susceptibilityprofiles}A). However, from our PFFRG calculations we find that for both oxides, and at all temperatures down till the RG flow breakdown, the maxima of the susceptibility are always located at the high-symmetry $W$ point in the extended Brillouin zone, i.e., at a $\mathbf{q}=\frac{2\pi}{a}(2,1,0)$-type ordering vector (see SI section S2), as found in Ref.~\onlinecite{Iqbal2019} for the nearest-neighbor isotropic $S=3/2$ Heisenberg antiferromagnet. A direct space schematic illustration for the spin configuration corresponding to this order is shown in Figure~\ref{fig:orders}A.   
In direct space, for this spin configuration, the pyrochlore lattice decomposes into two pairs of fcc sublattices. Within each pair, the two fcc sublattices are separated from each other by a nearest neighbor vector which is perpendicular to the $\frac{2\pi}{a}$ component of the ordering vector, e.g., for a $\frac{2\pi}{a}(2,1,0)$-ordering vector, one pair of fcc sublattices is separated by the vector $\frac{a}{4}(1,0,1)$ and the other pair is separated by the vector $\frac{a}{4}(1,0,-1)$. All fcc sublattices show a stripe antiferromagnetic order, i.e., spins are antiferromagnetically aligned in the $\frac{2\pi}{a}$-direction and ferromagnetically in the perpendicular directions. Within each pair, the neighboring spins of different sublattices are aligned antiparallel in spin space. The two pairs can be rotated freely with respect to each other. In reciprocal space, this configuration leads to main Bragg peaks at all $\frac{2\pi}{a}(2,1,0)$-type vectors which share the same $\frac{2\pi}{a}$ component, e.g., $\frac{2\pi}{a}(2,1,0)$, $\frac{2\pi}{a}(-2,1,0)$, $\frac{2\pi}{a}(0,1,2)$, $\frac{2\pi}{a}(0,1,-2)$, $\frac{2\pi}{a}(2,-1,0)$, $\frac{2\pi}{a}(-2,-1,0)$, $\frac{2\pi}{a}(0,-1,2)$, and $\frac{2\pi}{a}(0,-1,-2)$. In the reciprocal plane perpendicular to the $\frac{2\pi}{a}$-direction, the relative orientation of the sublattices results in subdominant Bragg peaks with half the intensity of the dominant ones at the $\frac{2\pi}{a}(1,0,1)$-type vectors in this plane, i.e., at $\frac{2\pi}{a}(1,0,1)$, $\frac{2\pi}{a}(-1,0,1)$, $\frac{2\pi}{a}(1,0,-1)$, $\frac{2\pi}{a}(-1,0,-1)$. For the classical model of {\lgo}, this state is the exact ground state given by the Luttinger-Tisza method and confirmed with the iterative minimization procedure. However, we find that this state is strongly dependent on the interplay between the two kinds of third nearest-neighbor couplings. This leads to the finding that for {\lio}, due to the larger difference between $J_{3a}$ and $J_{3b}$, an incommensurate order on the sublattices is stabilized as the ground state found by iterative minimization. We would like to emphasize that the stabilization of the orders discussed above rests crucially on the presence and the relative magnitude of the third nearest-neighbor couplings. Indeed, classically, the pure $J_1$-$J_2$-model on a regular $J_{1}\equiv J=J'$ pyrochlore lattice would feature a $\mathbf{q}=\mathbf{0}$ state, completely different from the $\frac{2\pi}{a}(2,1,0)$-order found here.

To make contact with experimental results, it is useful to assess the temperature dependence of the structure factor as calculated within PFFRG. The relation $T=(2\pi/3)S(S+1)\Lambda$ can be used to relate the infrared cut-off $\Lambda$ employed in the PFFRG framework to temperature $T$~\cite{Iqbal2019,Iqbal2016}. The relation is obtained by comparing the classical limit ($S\to\infty$) of PFFRG, where only the RPA diagrams contribute, i.e. a mean-field description, with the conventional spin mean-field theory formulated in terms of temperature $T$ instead of $\Lambda$. The resulting estimated ordering temperatures are given in SI section S2.  
As explained in detail in the Appendix A of Ref.~\onlinecite{Iqbal2019}, in the $S\to\infty$ limit, the absence of higher diagrammatic orders in $1/S$ in the one-loop PFFRG used here implies that, in particular for the nearest-neighbor pyrochlore antiferromagnetic Heisenberg model, a spurious divergence of the susceptibility at finite temperature occurs which is not expected from what is well established for the classical problem~\cite{Moessner1998,Iqbal2019}. Thus, our estimates of the ordering temperatures in all of the pyrochlore magnets considered here are systematically too high~\cite{Iqbal2018}.
Our comparisons to an experimental temperature $T$ are shown for a $\Lambda$ obeying $\Lambda/\Lambda_{\rm c}=T^{\rm exp}/T^{\rm exp}_{\rm c}$. It is expected that multi-loop implementations of PFFRG will lead to significant improvements for the calculated temperature scales~\cite{Rueck2018,Kugler2018}; however, these advanced schemes are still under development.

We have also performed the PFFRG calculation for the {\lio} Hamiltonian corresponding to its low temperature structure (see Table~\ref{tab:couplings}, last two lines). The calculated susceptibility profile (see SI Fig.~S4A-C) 
is seen to be virtually indistinguishable from that obtained from the room temperature structure shown in Figures~\ref{fig:susceptibilityprofiles}A-C even though the breathing anisotropy of the Hamiltonian is significantly different, having evolved from $J'/J=0.37$ at room temperature to $J'/J\cong 1$ at $T=20$~K. The general observation of the magnetic response being largely independent of the breathing anisotropy can be understood from a classical picture. As discussed in SI section S3, the eigenvalues of the interaction matrix $\mathcal{J}({\bf k})$ for a $J$-$J'$-only model exhibit flat bands at the lowest energies which determine the typical bow-tie features of pyrochlore antiferromagnets. Interestingly, these flat bands persist for any $J>0$ and $J'>0$~\cite{Benton2015} (see SI Fig.~S7A-D in SI section~S3), rendering the momentum space profile of the magnetic susceptibility independent of $J'/J$. As a consequence, longer range $J_2$, $J_{3a}$, $J_{3b}$ have the opportunity to play a crucial role in determining the magnetic behavior of the system. In the present case, however, we find that the effective second-neighbor coupling of {\lio} at 20 K given by $(J_2-J_{3a})/\bar{J}=-0.030$ is nearly unchanged compared to room temperature (see SI Table~S1) 
thus leading to almost identical magnetic responses.

It is now interesting to compare and verify the accuracy of the Hamiltonian determined for {\lio} against polarized neutron scattering data from powder samples of Okamoto {\it et al.}~\cite{Okamoto2015}. It was shown by Benton {\it et al.}~\cite{Benton2015} that within the self-consistent Gaussian approximation (SCGA), the experimental estimate $J'/J=0.1$, or even a range $0.05 \lesssim J'/J \lesssim 0.15$, is consistent with the neutron data. We also calculate the product $|F(Q)|^2S(Q)$ of the magnetic form factor $|F(Q)|^2$ and the powder averaged structure factor $S(Q)$ using 
  \begin{equation*}\begin{split}
      F(Q)&=-0.3094e^{-0.0274\big(\frac{Q}{4\pi}\big)^2}+0.36804e^{-17.0355\big(\frac{Q}{4\pi}\big)^2}\\&\mspace{20mu}+0.6559e^{-6.5236\big(\frac{Q}{4\pi}\big)^2}+0.2856\,,
\end{split}\end{equation*}
  for all the materials considered here~\cite{Brown2004}. The $S(Q)$ is calculated using both the quantum $S=3/2$ PFFRG and classical Monte Carlo (CMC) methods.  We show in Figures~\ref{fig:powderaverage}A-B the comparison to the $T=30$~K and the $T=150$~K neutron scattering data. The PFFRG calculations for the DFT determined Hamiltonians for {\lio} at room temperature ($J'/J=0.37$) as well as at $T=20$~K ($J'/J=1$), upon inclusion of the second- and third-neighbor couplings, give an excellent agreement with the $T=30$~K neutron data. In contrast, the CMC results, both for the room temperature as well as $T=20$~K couplings, progressively lose agreement as $Q$ is lowered. This observation may hint towards the non-negligible role of quantum fluctuations in accurately capturing the correlations especially at low $Q$ values. The excellent agreement of the earlier SCGA calculations of Ref.~\onlinecite{Benton2015} is also likely rooted in the fact that in this approach the ``hard constraint'', namely, that the spin vector on each site has magnitude $S$, is relaxed and only implemented on average. This may tacitly incorporate some effects of quantum fluctuations by allowing for fluctuations of the classical moments~\cite{Kimchi2014}. At $T=150$~K, the neutron data essentially follows the $Q$ dependence of the magnetic form factor~\cite{Okamoto2015}; while the computed scattering intensities for the room temperature and $T=20$~K Hamiltonians are consistent, the disagreement with the experimental $T=150$~K neutron data is neither satisfactory for the PFFRG nor for the CMC results. A similar observation has already been made by Benton {\it et al.}~\cite{Benton2015} and thus remains unexplained (see also SI section S2). 
\vspace{0.3cm}

\noindent{\bf Sulfides}

We have performed PFFRG calculations for all three chromium sulfides, {\lis}, {\lgs} and {\cis}. As we find very similar susceptibility profiles for the three compounds, we show the one for {\cis} in Figures~\ref{fig:susceptibilityprofiles}G-I and the other two in SI section S2, SI Figs.~S4D-I. {\cis} has a ferromagnetic $J'$ which is the strongest among all the interactions present in this material. At the low temperatures that we are considering here, we can therefore assume that the four spins on the large tetrahedra point in approximately the same direction, thereby behaving as effective spin-6 entities. Since the large tetrahedra are arranged in a face centered cubic (fcc) magnetic lattice (see SI Figs.~S5A-B), the system can effectively be mapped onto a spin-6 Heisenberg Hamiltonian on a fcc lattice with renormalized nearest-neighbor couplings $J_1^{\rm fcc}=(J+4J_2+2J_{3a}+2J_{3b})/16$. These couplings are all antiferromagnetic with values $J_1^{\rm fcc}=1.11$~K for {\lis}, $J_1^{\rm fcc}=0.95$~K for {\lgs} and $J_1^{\rm fcc}=2.56$~K for {\cis}. Interestingly, the nearest-neighbor antiferromagnetic classical Heisenberg model on the fcc lattice features a subextensive ($\mathcal{O}[L]$) ground state degeneracy in the classical limit ($S\rightarrow\infty$), with associated wave vectors of the form ${\bf q}=\frac{2\pi}{a}(1,\delta,0)$ and symmetry-related ${\bf q}$~\cite{Henley1987}. This degeneracy appears as lines of strong intensity in the magnetic response. However, as further explained in SI section~S5, the susceptibility profile of our pyrochlore model differs from that of the regular fcc antiferromagnet since it is modulated by a form factor arising from the presence of a tetrahedral basis of our effective fcc lattice. For the structure factors in the $[hk0]$ plane, this leads to a square ring type structure formed by lines of strong and almost constant intensity~\cite{Benton2015}, see Figure~\ref{fig:susceptibilityprofiles}H (and SI Figs.~S4E and H). The point ${\bf q}=\frac{2\pi}{a}(1,0,0)$ (and symmetry-related ${\bf q}$-space positions) is special because it resides at the junctions of the ground state manifold ${\bf q}=\frac{2\pi}{a}(1,\delta,0)$ and ${\bf q}=\frac{2\pi}{a}(1,0,\delta)$. Because of this, it is found that collinear ordered states with the ordering wave vector of the type $\mathbf{q}=\frac{2\pi}{a}(1,0,0)$ are selected by thermal~\cite{Henley1987} as well as quantum~\cite{Oguchi1985} {\it order-by-disorder} mechanism. As the PFFRG method incorporates the combined effects of thermal and quantum fluctuations, and thus of their selection effect, it is interesting to note that our analysis reveals maxima at ${\bf q}=\frac{2\pi}{a}(1,0,0)$ type positions for all three sulfides (see SI section S2). 
The corresponding direct space spin configuration is illustrated in Figure~\ref{fig:orders}B. The astonishing property that all sulfides show almost identical susceptibility profiles even though the ratio $J_1^{\rm eff}/J'$ varies significantly between the three compounds can be understood by considering the eigenvalues of the interaction matrix $\mathcal{J}({\bf k})$ for a pyrochlore system with $J>0$ and $J'<0$. As shown in SI Figs.~S7E-H of SI section S3, the lowest band exhibits line-like degeneracies, regardless of the size of $J/J'$, leading to a magnetic response which is largely independent of this ratio~\cite{Benton2015}.   

Plumier {\it et al.}~\cite{Plumier1971,Plumier1977} studied {\cis} using unpolarized neutrons. In order to compare our calculations to this experiment, we subtract the room temperature neutron spectra from the low temperature spectra in order to remove the background. In Figs.~\ref{fig:powderaverage}C-D, this data is compared to the calculated structure factor $S(Q)$ weighted with the magnetic form factor $|F(Q)|^2$ of Cr$^{3+}$ ions, i.e., $|F(Q)|^{2}S(Q)$. For the 40~K neutron data, the experiment matches well with our numerical data from PFFRG up to an arbitrary scale factor. At 85~K, while the peak positions match very well between theory and the experimental intensity profile, the intensity has some deviation at low $Q$. This could be partly due to the uncertainty in the mapping of the flow parameter $\Lambda$ to temperature. Since, 85~K is about 2.4$T_{\rm c}$ for {\cis}, the calculated structure factor is compared at 2.4$\Lambda_{\rm c}$. Nevertheless, the comparison of the spin-spin correlation profile to the available experimental information lends strong support to the validity of the Hamiltonian we determine for {\cis}. On the other hand, CMC calculations are unable to capture the dominant correlations present in the $40$~K experimental data, possibly hinting at the role of quantum fluctuations in accurately capturing the important correlations at low temperature, similar to what we observe in {\lio}.
\vspace{0.3cm}

\noindent{\bf Selenide}

The material {\cise} is the only selenide breathing chromium spinel with detailed structure reported for which we can perform calculations. At a first glance, the Hamiltonian parameters (Table~\ref{tab:couplings}) appear similar to those for {\lgs}, with a ferromagnetic large tetrahedron. However, the PFFRG result shown in Figure~\ref{fig:susceptibilityprofiles}J-L displays a completely different spin susceptibility. In contrast to a repetition of the broken high intensity lines representative of the one-dimensional sub-extensive degeneracy that we discussed in the previous subsection, we find the two-dimensional degeneracy of a sphere in reciprocal space. The ferromagnetic large tetrahedron characterized by $J'$ coupling form a large ($S=6$) moment fcc lattice with an effective nearest-neighbor interaction given by $J_1^{\rm fcc}=(J+4J_2+2J_{3a}+2J_{3b})/16$ which evaluates to $J_1^{\rm fcc}=-0.55$~K. The ferromagnetic nature of $J_1^{\rm fcc}$ in the selenide is indicative of the crucial difference between the selenide and the sulfides (for which the $J_1^{\rm fcc}$ are antiferromagnetic): in contrast to the sulfides, nearly negligible $J_2$ and smaller $J_{3a}$ and $J_{3b}$ couplings do not fully compensate the substantial ferromagnetic $J$ anymore. Nonetheless, clearly no simple ferromagnetic order is realized in {\cise}. Rather, the two third-neighbor couplings (taken together) substantially perturb the FM large tetrahedra which leads to the appearance of a two-dimensional degeneracy of classical ground states. At the ordering temperature (SI section S2), 
the system develops an incommensurate spiral magnetic order with a helix pitch vector $\mathbf{q}\approx\frac{2\pi}{a}(0.521,0,0)$ (and symmetry-related points $(0,q,0)$ and $(0,0,q)$). The corresponding magnetic susceptibility profile obtained from PFFRG at the ordering temperature is shown in Figures~\ref{fig:susceptibilityprofiles}J-K wherein one observes peaks in $S(\mathbf{q})$ at $\mathbf{q}\neq\mathbf{0}$. The corresponding direct space spin configuration is illustrated in Figure~\ref{fig:orders}C. It is worth emphasizing that the shift of the spectral weight away from $\mathbf{q}=\mathbf{0}$ is the combined effect of \emph{both} $J_{3a}$ and $J_{3b}$ couplings of a comparable magnitude~\cite{Tymoshenko2017}. Above the ordering temperature, and within the cooperative paramagnetic regime, we observe that the spectral weight is distributed rather uniformly over the spiral surface suggesting the presence of an approximate spiral spin liquid on the pyrochlore lattice stabilized concomitantly by thermal and quantum fluctuations. While an incommensurate long-range ordered spiral phase of $(q,0,0)$ type has been observed for the material \ce{ZnCr2Se4} with pyrochlore magnetic lattice~\cite{Tymoshenko2017}, it is only very recently that a spiral spin liquid has been observed in the chromium spinel MgCr$_{2}$O$_{4}$~\cite{Bai2019}.

The spin spiral surface found in the cooperative paramagnetic regime in a PFFRG calculation, is also found to be present in the corresponding classical model at $T=0$, as revealed by a Luttinger-Tisza analysis. However, the Luttinger-Tisza eigenstates on this spiral surface generically \emph{do not} represent real normalizable spin configurations, i.e., the spin length on different fcc sublattices are not equal in magnitude. It turns out that, it is only for the $(q,0,0)$-type ordering vectors (with $q\approx \frac{2\pi}{a}(0.45,0,0)$) that the minimal energy can be achieved by a configuration of normalized spins. This ground state configuration then breaks the symmetry of the lattice as it is only governed by one of the three $(q,0,0)$-type vectors. Indeed, our classical  Monte Carlo calculations give an  ordering vector of $\mathbf{q}=\frac{2\pi}{a}(0.40\pm0.04)$ at the transition temperature. Although the spiral surface is inaccessible at $T=0$ due to the violation of the spin length constraint, thanks to thermal and/or quantum fluctuations it can in principle be made accessible. Indeed, for {\cise}, our PFFRG analysis shows that thermal and quantum fluctuations taken together are able to restore a well-defined spin spiral surface (see Figs.~\ref{fig:susceptibilityprofiles}J-K) right above the ordering temperature. In contrast, our classical Monte Carlo simulations, right above the ordering temperature, find a highly nonuniform distribution of spectral weight, and thus an absence of spiral spin liquid, however, as the temperature is increased to $\sim1.5$ times the ordering temperature, a spiral surface appears (compare SI Fig.~S8 of SI section~S4 with Figs.~\ref{fig:susceptibilityprofiles}J-K above). These findings lend support to the role played by quantum fluctuations in aiding the stabilization of a spiral spin liquid. It is worth noting that within the ordered phase, this single-$\mathbf{q}$ state does not however represent a single planar spiral throughout the entire lattice. Instead, similar to what is found in {\cis}, the lattice can be divided into two pairs of fcc sublattices. Within each pair, the sublattices are connected by a nearest neighbor vector perpendicular to the ordering vector, and the planar spiral orders within each pair are of equal pitch and phase. However, the two pairs of fcc sublattices are found to be out of phase with respect to each other by approximately $5^\circ$. This offset in phase is a direct and unique consequence of the breathing anisotropy, and is found to vanish in the isotropic pyrochlore lattice.
\vspace{0.3cm}

\noindent{\bf Discussion}

One of the important findings of this study is the significant and heretofore unappreciated role played by long-range exchange interactions on the breathing pyrochlore lattice.  While we have previously pointed out the sensitivity of the spin structure factor to the sign and size of the subleading couplings in the $J_1$-$J_2$ quantum pyrochlore Heisenberg antiferromagnet~\cite{Iqbal2019}, we have found in the present work for the chromium breathing pyrochlores an impressive demonstration of the decisive role played by longer-range couplings, i.e., the two kinds of symmetry inequivalent third-nearest neighbor couplings $J_{3a}$ and $J_{3b}$. On the other hand, the effects of breathing anisotropy are shown to be surprisingly minor. Indeed, we observe that in {\lio}, the effective ferromagnetic second-neighbor coupling of 2.5{\%} of the average nearest-neighbor coupling far outweighs the substantial room temperature breathing anisotropy of $J'/J=0.37$ in its effect on the structure factor profile. In the case of the sulfides, although the $J'/J$ ratios vary between $93$ and $-1.8$ among the three compounds, the susceptibility profiles appear all very similar. In this case, this is rooted in the effective mapping of the system to an emerging effective nearest-neighbor fcc Heisenberg antiferromagnet with the fcc lattice sites being occupied by ferromagnetic tetrahedra featuring a large $S=6$ magnetic moment. In this case, we show that the spin structure factor displays line-like degeneracies which are, once more, largely independent of the breathing anisotropy ratio $J'/J$. Finally, for the dominantly ferromagnetic selenide {\cise}, we demonstrate that the combined effect of the two third-neighbor couplings $J_{3a}$ and $J_{3b}$ drastically perturbs the ground state away from the simple ferromagnetic order. Interestingly, we find that the perturbed state corresponds to an approximate spiral spin liquid above the ordering temperature and within the cooperative paramagnetic regime, where the individual wave vectors form a sphere-like manifold in reciprocal space. Our PFFRG analysis indicates that the combined effect of quantum and thermal fluctuations only leads to a weak order-by-disorder selection into an incommensurate spiral state. Approximate spiral spin liquids on three-dimensional lattices are so far known only on the diamond lattice in MnSc$_{2}$S$_{4}$~\cite{Bergman2007,Gao2016,Iqbal2018}, with a very recent occurrence having also been reported for the pyrochlore material MgCr$_{2}$O$_{4}$~\cite{Bai2019}. 
\vspace{0.3cm}

\noindent{\bf Conclusion}

We have theoretically investigated six chromium spinels featuring crystalline, and thus magnetic exchange breathing anisotropy. The Hamiltonians which were determined from density functional theory calculations and investigated using the pseudofermion functional renormalization group method showcase a colorful selection of magnetic properties arising from frustrated interactions. We find that the oxide compounds are in a perturbed Coulomb phase~\cite{Moessner1998}, reminiscent of the famous spin ice materials. The sulfides display an effective fcc lattice Heisenberg antiferromagnetic type behavior, and the selenide features incommensurate magnetic correlations. As a unifying feature, all materials are found to be close to a classical degeneracy which shows up as different variants: The oxides are near a phase featuring an extensive number of classical ground states scaling exponentially in the volume of the system. The selenide is approximately degenerate on a two-dimensional surface with a manifold of states scaling exponentially in $L^{2}$ (where $L$ is the linear dimension of the system). Finally, the sulfides are characterized by approximate line-like degeneracies with the number of ground states scaling exponentially in the linear dimension $L$. This variety of different behaviors in a family of related materials represents an attractive feature which promises a wealth of opportunity for future investigations; on the one hand, doping series interpolating between different types of ground states could be very interesting. On the other, it is an intriguing question how other known but only sketchily characterized breathing chromium spinels like \ce{CuGaCr4S8}, \ce{AgInCr4S8}, \ce{CuGaCr4Se8}, \ce{AgInCr4Se8}~\cite{Haeuseler1977} fit the picture outlined in this study. Of particular importance are experimental investigations with single crystals which would allow to assess our theoretical predictions. 
\vspace{0.3cm}

\noindent{\bf Materials and Methods}

\noindent{\bf Energy mapping method}

We determine the electronic structure and total energies of the breathing pyrochlores using the full potential local orbital (FPLO) basis set~\cite{Koepernik1999} and the generalized gradient approximation (GGA) functional~\cite{Perdew1996}. The GGA+$U$~\cite{Liechtenstein1995} correction to the exchange and correlation functional is used to deal with the strong electronic correlations on the Cr$^{3+}$ $3d$ orbitals. Two parameters enter these calculations, the on-site interaction strength $U$ and the Hund's rule coupling $J_{\rm H}$. As the Hund’s rule coupling is an intra-atomic interaction, we do not expect it to vary much from one compound to the other, and thus henceforth fix its value to $J_{\rm H} = 0.72$ eV as commonly employed for Cr$^{3+}$~\cite{Mizokawa1996}. Meanwhile, the on-site interaction $U$ was fitted from the experimentally determined Curie-Weiss temperatures as explained below. Heisenberg Hamiltonian parameters are extracted using the energy mapping approach~\cite{Jeschke2011,Iqbal2018}: For this purpose, a $3\times 1\times 1$ supercell of the primitive unit of the cubic $F\bar{4}3m$ structure with $Cm$ space group is constructed. In this structure, there are 9 independent spins (out of 12 in total), allowing 75 spin configurations with different energies. Total energies are converged using $6\times 6\times 6$ $\mathbf{k}$ points. An arbitrary subset of 19 spin configurations allows us to fit six exchange interactions, using a Heisenberg Hamiltonian of the form  of Eq.~\eqref{eq:H}. Note that we count every bond only once. The quality of the fit is seen to be excellent (see SI Fig.~S2 in SI section~S1 for an example), indicating that the method works extremely well for the chromium breathing pyrochlores, may they be dominantly antiferromagnetic or ferromagnetic. Now, the $U$ parameter of the GGA+$U$ functional can be fixed: We calculate the Hamiltonian parameters for a small set of $U$ values, which we choose such that the Curie-Weiss temperatures $\uptheta_{\rm CW}=-\frac{1}{3}S(S+1)(3J+3J'+12J_{2}+6J_{3a}+6J_{3b})$ (see Figure~\ref{fig:paths}) calculated from these Hamiltonians cover a range of temperatures which includes the experimental $\uptheta_{\rm CW}$. We find the $U$ value from the condition $\uptheta^{\rm theor}_{\rm CW}=\uptheta^{\rm exp}_{\rm CW}$ by interpolation (see Figure~\ref{fig:couplings}). For the breathing pyrochlores, we find the $\uptheta^{\rm theor}_{\rm CW}$ to be monotonous functions of $U$ and thus to allow for a unique solution. Some more discussion on $\uptheta_{\rm CW}$ and $U$ can be found in SI section S1.
\vspace{0.3cm}

\noindent{\bf Pseudofermion functional renormalization group method}

In the PFFRG scheme~\cite{Reuther2010}, a general spin-$S$ Heisenberg Hamiltonian, e.g. Eq.~(\ref{eq:H}), is treated by first re-expressing  the spin operator at each lattice site by $2S$ spin-1/2 degrees of freedom using Abrikosov pseudofermions. Aside from the desired Hilbert space of spin-$S$ states this approach also introduces unphysical states with lower spin magnitudes. Since these states contribute less to the energy of the system compared to the physical ones, they act as excitations and, thus, naturally get excluded at low temperatures in the RG process~\cite{Iqbal2019}. Next, the pseudo-fermionic Hamiltonian is treated using the Functional Renormalization Group (FRG) formalism~\cite{Reuther2010,Metzner2012,Baez2017,Buessen2018}. Here, the bare fermionic Green's function in Matsubara space is equipped with an infrared frequency cut-off $\Lambda$ suppressing the fermionic propagation at low frequencies. The PFFRG flow equations then follow from the $\Lambda$ dependence of the $m$-particle irreducible vertex functions which are given by an exact and infinite hierarchy of coupled integro-differential flow equations~\cite{Metzner2012}. For a numerical implementation, this hierarchy however needs to be truncated, which is done by neglecting three-particle and higher vertex functions. Still, and most importantly, the truncated version of the method exactly sums up the full set of fermionic Feynman diagrams in two limits taken separately, namely, the large-$S$ limit~\cite{Baez2017} and the large-$N$ limit~\cite{Buessen2018}. In doing so, it provides an unbiased analysis of the competition between magnetic ordering and disordering tendencies~\cite{Iqbal2019,Reuther2010,Metzner2012,Baez2017}. 

The flow equations are numerically solved in direct space and, after the Fourier transform of the spin-spin correlations, one obtains the static susceptibility in reciprocal space $S(\mathbf{q})$ as a function of $\Lambda$. Note that the RG parameter $\Lambda$ can be interpreted as the temperature $T$~\cite{Iqbal2019} and is therefore used as a proxy to investigate the thermal evolution of the magnetic response. In the present study, we evaluate spin correlators within a cube of edge length of $5a$, where $a$ is the cubic lattice constant, which incorporates a total of 2315 correlated sites, producing well-converged results with a high $\mathbf{q}$-space resolution. Furthermore, we approximate the frequency dependence of the vertex functions by discrete grids containing 64 points for each frequency variable. An ordered system is identified by cusps or kinks in the susceptibility flow signaling a magnetic instability towards either magnetic or valence bond order~\cite{Reuther2010}. In contrast, a smooth flow of the susceptibility down to $\Lambda\rightarrow 0$ indicates a magnetically disordered state, that is, a putative (quantum) spin liquid~\cite{Reuther2010}.
\vspace{0.3cm}

\noindent{\bf Luttinger-Tisza method}

In the Luttinger-Tisza method~\cite{Kaplan2007,Lapa2012} the ground state and energy spectrum of a classical spin system are approximately calculated. The classical limit of Eq.~(\ref{eq:H}) consists of replacing the spin operators by classical vectors, which are normalized (the so called \emph{strong constraint}). To find an approximate ground state, this constraint is relaxed to be fulfilled only on average over the whole system. This allows to Fourier transform the interaction matrix $J_{ij}$ with respect to the underlying Bravais lattice of the actual spin lattice, leading to a matrix $J_{\alpha\beta}(\mathbf{k})$, where $\alpha, \beta$ label the basis points of the lattice.

The eigenvalues $\lambda(\mathbf{k})$ of this matrix subsequently give the energies of spin spiral states with wave vector $\mathbf{k}$, the aforementioned energy spectrum, and the absolute value of the corresponding eigenvector components give the relative length of the vector spins on the different basis points. This means, that for such a state to fulfill the strong constraint, all components have to have equal absolute value.

If the lattice is of Bravais type, there is only one basis point and therefore the Fourier transformed interaction matrix is a scalar, which means this condition is trivially fulfilled, rendering the Luttinger-Tisza method exact on such lattices. In this case, we can also, through the equivalence of the Luttinger-Tisza method and PFFRG for $S\to\infty$~\cite{Baez2017}, calculate the classical spin susceptibility
\begin{equation}
\chi(\mathbf{k}) \sim \frac{1}{\frac{2}{3 T}-\frac{1}{J(\mathbf{k})}}.
\end{equation}

\noindent{\bf Iterative minimization method}

By employing the iterative minimization method~\cite{Lapa2012,Iqbal2019}, we find the ground state of the classical version of Eq.~\eqref{eq:H} obtained by replacing the spin operators by normalized vectors. In this scheme, we typically start from a random spin configuration, randomly select one vector-spin $\mathbf{S}_i$ at a time and rotate it to make it antiparallel to its local field,
\begin{equation}
\mathbf{h}_i \equiv \frac{\delta H }{\delta \mathbf{S}_i}\,,
\end{equation}
thus minimizing the energy. In a given sweep, we update all the spins contained in the configuration so that on average each spin is updated once. For our calculations, we use 32 cubic unit cells of the pyrochlore lattice in each direction with periodic boundary conditions, totaling to $16\times 32^3=\num{524288}$ spins in the system. Convergence is reached once the total energy difference per sweep is less than $10^{-10}$~K.  To reduce the likelihood of the system getting stuck at a local energy minimum in configuration space, we carry out the minimization for at least 20 different random initial configurations. We also use specific non-random configurations, e.g. spin spirals, as a starting point, to test exact spin parameterizations and, possibly, find lower minimal energies than those achieved by starting with arbitrary random spin configurations.
\vspace{0.3cm}

\noindent{\bf Classical Monte Carlo method}

We have performed classical Monte Carlo simulations for the model Hamiltonian parameters of the compounds as obtained from DFT given in Table~\ref{tab:couplings}, employing the standard single spin-flip technique. For each Monte Carlo update, we perform the Metropolis moves as many times as the number of spins in the configuration such that, on average, each spin is updated once. Starting from a random spin configuration and after allowing for the system to reach thermal equilibrium, we evaluate the Fourier transform of the equal-time spin-spin correlator, i.e., the static structure factor $S(\mathbf{q})$ for 50 different configurations where each configuration is separated by $100$ Monte Carlo updates. At lower temperatures, in order to increase the number of accepted moves, we restrict the newly generated spins to a small cone around its predecessor. The simulations are done for a system of $32\times32\times32$ cubic unit cells of the pyrochlore lattice with periodic boundary conditions in each direction, amounting to $16 \times 32^3$ spins. The entire simulations are independently carried out 15 times and the average of the results is taken for suppression of numerical noise.
\vspace{0.3cm}


\vspace{0.3cm}

\noindent{\bf Acknowledgments:}
We thank Zenji Hiroi, G{\o}ran Nilsen, Yoshihiko Okamoto, Owen Benton, Hikaru Kawamura and Jason Gardner for useful discussions. H. O. J. thanks Gøran Nilsen for communicating the $T=20$~K internal coordinates of {\lio}. We gratefully acknowledge the Gauss Centre for Supercomputing e.V. for funding this project by providing computing time on the GCS Supercomputer SuperMUC at Leibniz Supercomputing Centre (LRZ).
{\bf Funding:} The work in W\"urzburg is funded by the Deutsche Forschungsgemeinschaft (DFG, German Research Foundation)-Project-ID 258499086-SFB 1170 and by the W\"urzburg-Dresden Cluster of Excellence on Complexity and Topology in Quantum Matter -- ct.qmat (EXC 2147, project-id 39085490). This research was supported in part by the International Centre for Theoretical Sciences (ICTS) during a visit for participating in the program - The 2nd Asia Pacific Workshop on Quantum Magnetism (Code: ICTS/apfm2018/11). J.R. acknowledges the kind hospitality of the Indian Institute of Technology Madras, Chennai, India where a part of the research was carried out. The work at the University of Waterloo was supported by the Canada Research Chair program (M.J.P.G., Tier 1).

{\bf Author contributions} H.O.J. initiated the project. P.G., Y.I., T.M., J.R., R.T.P. and H.O.J. performed the calculations. All authors analyzed the data and contributed to their interpretation. P.G., Y.I., M.J.P.G. and H.O.J. wrote the manuscript with input from all authors.

{\bf Competing interests:} The authors declare that they have no competing interests.

{\bf Data and materials availability:} All data needed to evaluate the conclusions in the paper are present in the paper and/or the Supplementary Materials. Additional data related to this paper may be requested from the authors.


\clearpage


\section*{Supplementary material}

# Breathing chromium spinels: a showcase for a variety of pyrochlore Heisenberg Hamiltonians
## – Supplementary Information –

Pratyay Ghosh,[1] Yasir Iqbal,[1] Tobias Müller,[2] Ravi T. Ponnaganti,[1] Ronny Thomale,[2]
Rajesh Narayanan,[1] Johannes Reuther,[3] Michel J. P. Gingras,[4,5] and Harald O. Jeschke[6]

[1] Department of Physics, Indian Institute of Technology Madras, Chennai 600036, India
[2] Institute for Theoretical Physics and Astrophysics,
Julius-Maximilians-Universität Würzburg, Am Hubland, 97074 Würzburg, Germany
[3] Dahlem Center for Complex Quantum Systems and Fachbereich Physik,
Freie Universität Berlin, 14195 Berlin, Germany
[4] Department of Physics and Astronomy, University of Waterloo, Waterloo, Ontario, N2L 3G1, Canada
[5] Quantum Materials Program, Canadian Institute for Advanced Research, MaRS Centre,
West Tower 661 University Ave., Suite 505, Toronto, ON, M5G 1M1, Canada
[6] Research Institute for Interdisciplinary Science, Okayama University, Okayama 700-8530, Japan

(Dated: October 16, 2019)

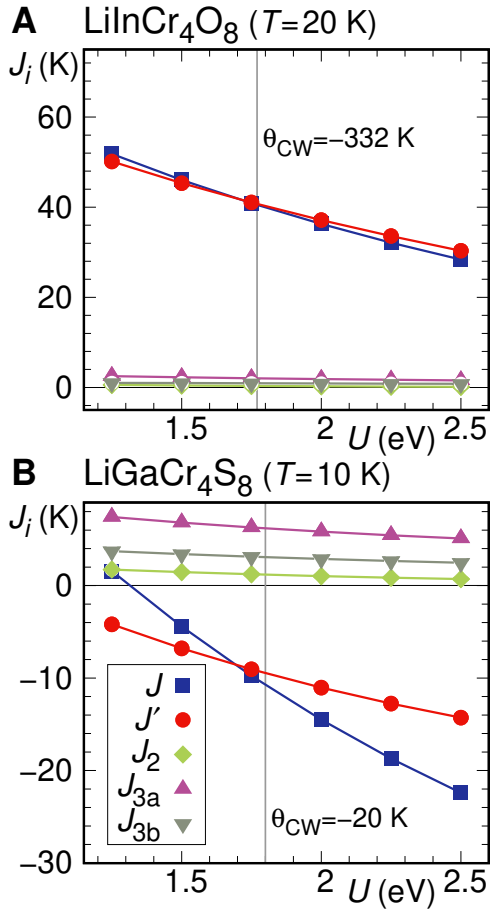

FIG. S1. **Exchange couplings of two breathing chromium spinels.** The values are calculated by mapping of DFT total energies for low temperature structures of (**A**) LiInCr$_4$O$_8$ (9) and (**B**) LiGaCr$_4$S$_8$ (18). Couplings are shown as function of interaction strength $U$. Vertical lines indicate the $U$ value at which the experimental Curie-Weiss temperature is realized. The $T = 20$ K internal coordinates of LiInCr$_4$O$_8$ were privately communicated by Goran Nilsen.

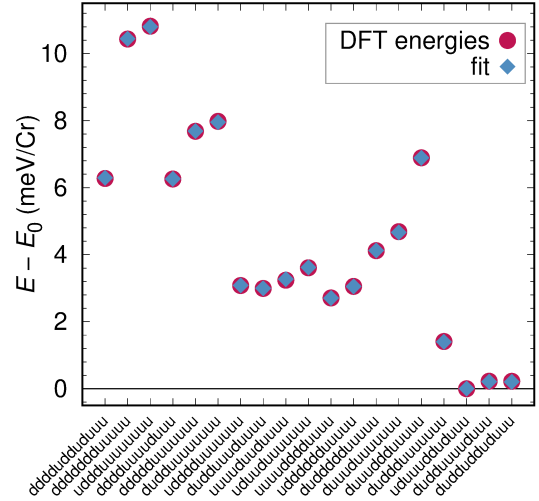

FIG. S2. **Quality of the energy mapping scheme.** Density functional theory energies for 19 different spin configurations of LiInCr$_4$O$_8$ calculated with GGA+$U$ exchange correlation functional at $U = 2$ eV and $J_{\rm H} = 0.72$ eV. The fit to the Heisenberg Hamiltonian with six exchange couplings is excellent.

## S1. ADDITIONAL DFT DETAILS

We show the result of the energy mapping for the two low temperature structures of LiInCr$_4$O$_8$ and LiGaCr$_4$S$_8$ in fig. S1.

The energy mapping scheme (55,56) for the chromium breathing pyrochlores works very well. Total moments are exact multiples of 3 $\mu_{\rm B}$ as all chromium moments are exactly $S = 3/2$. All fits are extremely good, with an example given in fig. S2.

For convenience, we show in table S1 the average nearest-neighbor energy scale $\bar{J} = \sqrt{J^2 + J'^2}$ for all compounds and all (room temperature and low temperature) structures on record as discussed in the main text (see Table 1).

TABLE S1. **Overview of subleading exchange couplings.** Longer range exchange couplings of the chromium breathing pyrochlore compounds are given relative to the average nearest-neighbor energy scale $\bar{J} = \sqrt{J^2 + J'^2}$. The absolute numbers are provided in Table 1 of the main text.

| Material | $T$ (K) | $\bar{J}$ (K) | $J_2/\bar{J}$ | $J_{3a}/\bar{J}$ | $J_{3b}/\bar{J}$ | $(J_2 - J_{3a})/\bar{J}$ |
|---|---|---|---|---|---|---|
| LiInCr$_4$O$_8$ | 20 | 57.5 | 0.006 | 0.035 | 0.016 | −0.030 |
| LiInCr$_4$O$_8$ | RT | 63.7 | 0.004 | 0.029 | 0.015 | −0.025 |
| LiGaCr$_4$O$_8$ | RT | 120.0 | 0.006 | 0.016 | 0.011 | −0.011 |
| LiInCr$_4$S$_8$ | RT | 27.8 | 0.025 | 0.190 | 0.086 | −0.165 |
| LiGaCr$_4$S$_8$ | 10 | 14.3 | 0.083 | 0.432 | 0.215 | −0.350 |
| LiGaCr$_4$S$_8$ | RT | 14.4 | 0.081 | 0.422 | 0.210 | −0.341 |
| CuInCr$_4$S$_8$ | RT | 29.6 | 0.038 | 0.215 | 0.153 | −0.177 |
| CuInCr$_4$Se$_8$ | RT | 40.1 | 0.008 | 0.117 | 0.094 | −0.109 |

TABLE S2. **Structural breathing anisotropy.** The two smallest Cr$^{3+}$–Cr$^{3+}$ distances $r$ and $r'$ for the investigated breathing pyrochlores are listed together with the (lattice) breathing anisotropy $r'/r$. RT stands for room temperature.

| Material | $T$ (K) | $r \equiv d^J_{\mathrm{Cr-Cr}}$ (Å) | $r' \equiv d^{J'}_{\mathrm{Cr-Cr}}$ (Å) | $r'/r$ |
|---|---|---|---|---|
| LiInCr$_4$O$_8$ (9) | 20 | 2.929 | 3.014 | 1.029 |
| LiInCr$_4$O$_8$ (6) | RT | 2.903 | 3.051 | 1.051 |
| LiGaCr$_4$O$_8$ (6) | RT | 2.867 | 2.970 | 1.036 |
| LiInCr$_4$S$_8$ (16) | RT | 3.429 | 3.735 | 1.089 |
| LiGaCr$_4$S$_8$ (15) | 10 | 3.404 | 3.641 | 1.070 |
| LiGaCr$_4$S$_8$ (16) | RT | 3.397 | 3.649 | 1.074 |
| CuInCr$_4$S$_8$ (16) | RT | 3.405 | 3.709 | 1.089 |
| CuInCr$_4$Se$_8$ (7) | RT | 3.613 | 3.864 | 1.070 |

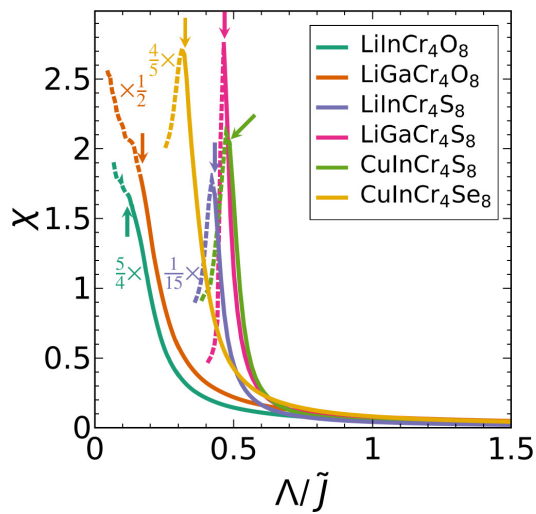

FIG. S3. **Detection of magnetic order.** RG flows of the maximum of magnetic susceptibilities of all the six compounds under investigation. The points at which the solid lines become dashed indicate an instability in the flow and express the onset of a magnetic order.

It is an important finding that the breathing anisotropy found in the exchange couplings is not simply connected to the structural breathing anisotropy. To facilitate the comparison, we list in table S2 the two Cr-Cr distances within small and large tetrahedra for the six room temperature and the two low temperature structures.

It is known that the paramagnetic Curie law in pyrochlores may only be reached at rather high temperatures, leading to possible inaccuracies in experimentally determined Curie-Weiss temperatures (57) In the case of the breathing pyrochlores, the Curie-Weiss fits were done in a 200 K to 350 K range for LiInCr$_4$O$_8$ and LiGaCr$_4$O$_8$ (6), in a 200 K to 300 K range for LiInCr$_4$S$_8$, LiGaCr$_4$S$_8$ and CuInCr$_4$S$_8$ (16) and using temperatures up to 380 K for CuInCr$_4$Se$_8$ (58) Note that the conclusions of this work will not change even if a 10% or 20% underestimation of the Curie-Weiss temperatures would force us to revise down the $U$ values at which we read off the relevant model Hamiltonians in Figure 2 of the main text. For the oxides, the Hamiltonian will remain in the same limit even if $U$ is lowered substantially. The same is true for the sulfides because the FM coupling of the large tetrahedron does not depend much on $U$.

## S2. ADDITIONAL PFFRG DATA

In the pseudo-Fermion functional renormalization group (PFFRG) approach, the onset of magnetic long-range order is signaled by a breakdown of the smooth renormalization group (RG) flow. Figure S3 shows the flows computed for the room temperature structures of all six breathing pyrochlore compounds considered in this study. The continuation of the flow below the breakdown energy scale is shown with a dashed line as the PFFRG calculation becomes less reliable in the ordered state. The ordering temperatures determined in this way are the following: LiInCr$_4$O$_8$ orders at 60.6 K with ordering wave vector $\mathbf{q} = (2,1,0)$ (experiment $T_N = 15.9$ K (6). LiGaCr$_4$O$_8$ orders at 159.8 K with ordering wave vector $\mathbf{q} = (2,1,0)$ (experiment $T_N = 13.8$ K (6). LiInCr$_4$S$_8$ orders at 95.0 K with ordering wave vector $\mathbf{q} = (1,0,0)$ (experiment $T_c = 24$ K (16). LiGaCr$_4$S$_8$ orders at 49.1 K with ordering wave vector $\mathbf{q} = (1,0,0)$ (experiment $T_c = 10$ K (15,16). CuInCr$_4$S$_8$ orders at 114.7 K with ordering wave vector $\mathbf{q} = (1,0,0)$ (experiment $T_c = 40$ K (38), $T_c = 35$ K (39), 32 (16). CuInCr$_4$Se$_8$ orders at 99.3 K

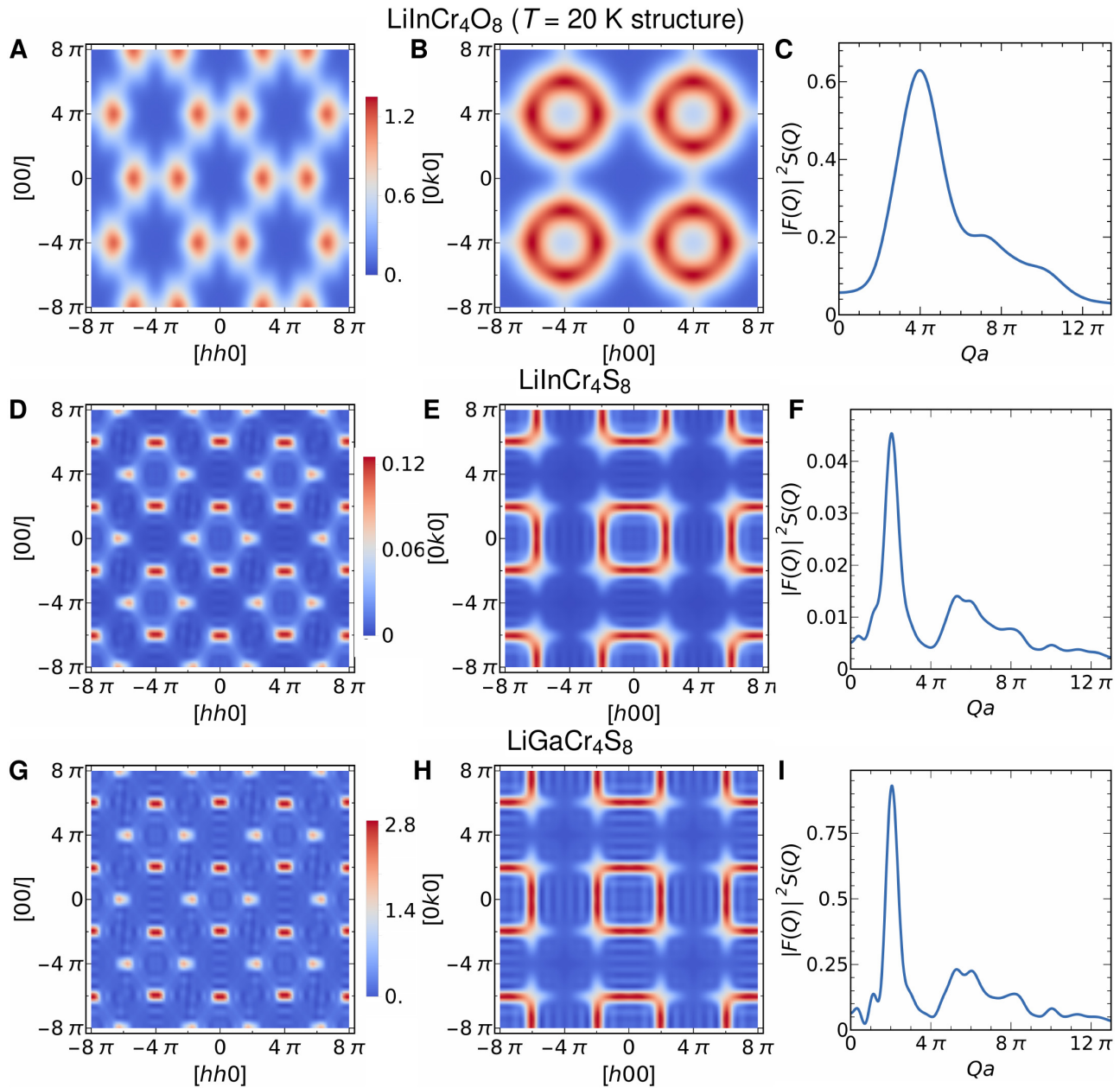

FIG. S4. **Additional spin susceptibility profiles.** PFFRG results for the $T = 20$ K Hamiltonian of LiInCr$_4$O$_8$, for LiInCr$_4$S$_8$ and for LiGaCr$_4$S$_8$ are shown. They are evaluated at the breakdown point in the RG flow, marked by an arrow in fig. S3. (**A**, **D**, **G**) Spin susceptibility projected on the $[hhl]$ plane. (**B**, **E**, **H**) Spin susceptibility projected on the $[hk0]$ plane. Susceptibility units are $1/\bar{J}$ ($\bar{J} = \sqrt{J^2 + J'^2}$), and the axes are in units of inverse lattice parameter $1/a$. (**C**, **F**, **I**) Plots of the form factor modulated powder averaged susceptibility $|F(Q)|^2 S(Q)$ vs. $Qa$.

with ordering wave vector $\mathbf{q} = (0.521, 0, 0)$ (experiment $T_c = 14$ K (59).

**Oxides.-** The room temperature susceptibility profiles of LiInCr$_4$O$_8$ is shown in the main text (see Figures 3A-B). The PFFRG calculation with the low temperature structure of LiInCr$_4$O$_8$ yields the susceptibility profiles shown in figure S4A-B which is calculated with the $T = 20$ K Hamiltonian parameters from Table 1 of the main text. Even though the breathing anisotropy is reduced from $J'/J = 0.37$ to $J'/J = 1$, the two static spin susceptibility profiles are very similar.

Concerning the disagreement of both PFFRG and classical Monte Carlo (18) calculations with the experimental $T = 150$ K neutron data, we speculate that a possible suppression of magnetic scattering at low angles due to the strong neutron absorption of In might explain this discrepancy and would be worth further investigating, for example by performing the same experiment on

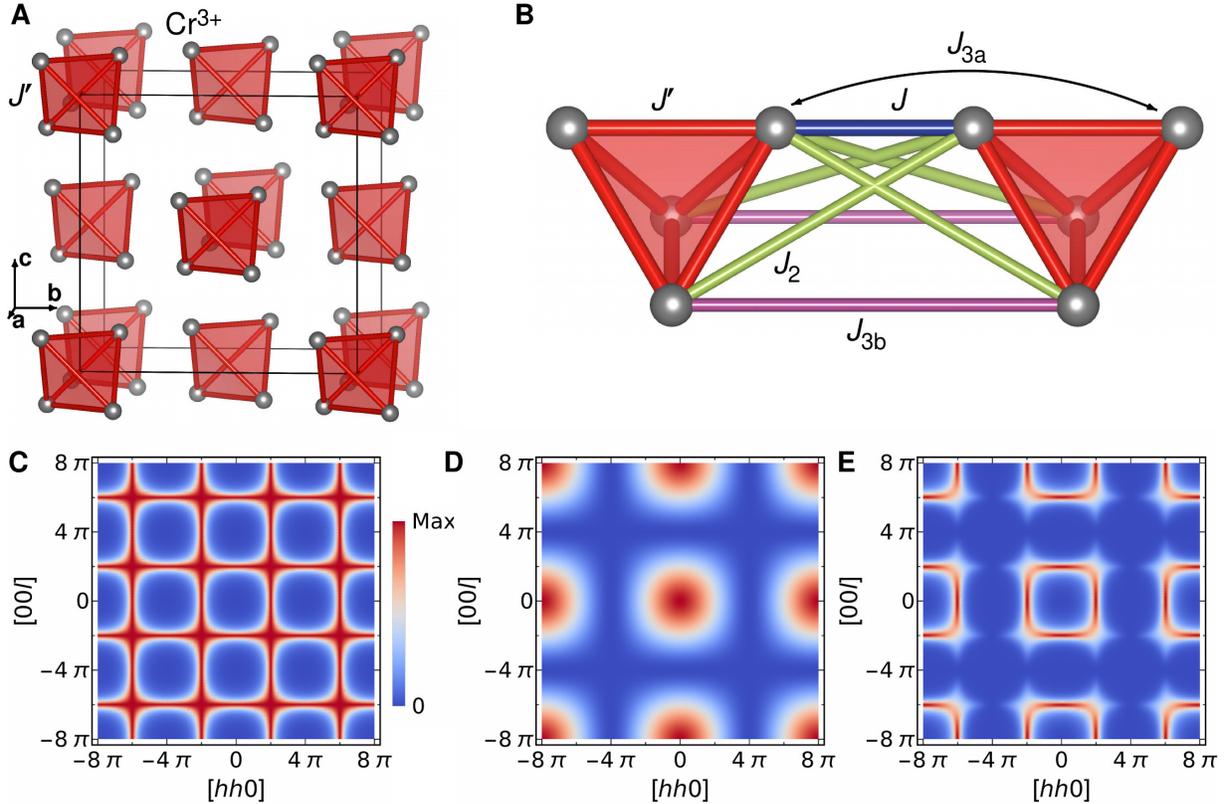

FIG. S5. **Consequences of the effective fcc lattice.** (**A**) Large tetrahedra of the chromium sulfides, forming an fcc lattice. Only the strong ferromagnetic $J'$ coupling is depicted which turns the $Cr^{3+}$ tetrahedra into approximate $S = 6$ units. (**B**) Exchange interactions between nearest-neighbor large tetrahedra in the fcc structure. (**C**) Classical spin susceptibility of the fcc nearest-neighbor Heisenberg antiferromagnet (NNAF) just above the ordering temperature; the critical temperature of the NNAF on the fcc lattice is $T_c = \frac{4}{3}J$, the plot is done at $T = 1.48J$. (**D**) Structure factor $|f(\mathbf{q})|^2$ of the pyrochlore basis. (**E**) Classical spin susceptibility of a effective fcc nearest-neighbor model on the pyrochlore lattice formed by ferromagnetically ordered tetrahedra in (A), given by the product of (C) and (D).

$LiGaCr_4O_8$ (Ga has a 70 times smaller neutron absorption cross section than In (60). Another possible explanation involves the inelasticity of scattering at high temperatures which could explain a shift of spectral weight to low Q.

**Sulfides.-** The spin susceptibility profile of $LiGaCr_4S_8$ is shown in the main text (see Figure 3G-I). The other two sulfide spin susceptibility profiles are shown in figs S4D-I. All three sulfide PFFRG results are rather similar even though the Hamiltonians do not appear so similar at the outset. As explained in the main text, this can be understood by recognizing that the ferromagnetic large tetrahedra form an fcc lattice of large $S = 6$ magnetic moments, as illustrated in fig. S5A. The renormalized nearest-neighbor interaction in the fcc lattice is given by $J_1^{fcc} = (J + 4J_2 + 2J_{3a} + 2J_{3b})/16$ as can be read off from fig. S5B. For the three sulfides, $J_1^{fcc}$ turns out to be antiferromagnetic.

In the main text, we show that the form factor modulated powder averaged susceptibility $|F(Q)|^2 S(Q)$ which we calculated for $CuInCr_4S_8$ matches the experimental data of Refs. 38 and 39 very well, accurately reproducing three peak positions at $T = 85$ K (see Figure 5B of the main text). We wish to investigate the question to what extent the subleading exchange couplings are important for this very good agreement. To do so, we calculate the $|F(Q)|^2 S(Q)$ for Hamiltonians where we set selected longer range couplings to zero; we consider the set $J$-$J'$-$J_2$ ($J_{3a}$ and $J_{3b}$ are neglected), the set $J$-$J'$-$J_2$-$J_{3a}$ ($J_{3b}$ is neglected) and the set $J$-$J'$-$J_2$-$J_{3b}$ ($J_{3a}$ is neglected). The results are shown in figure S6. From both plots, it's clear that the best match with the experimental data is obtained only with the full $J$-$J'$-$J_2$-$J_{3a}$-$J_{3b}$ model. Each omission of a subleading coupling leads to a worse match with experiment. This validates the values we determined for $CuInCr_4S_8$ using DFT energy mapping and underlines the crucial role played by the subleading couplings up to third nearest neighbor in the breathing pyrochlores.

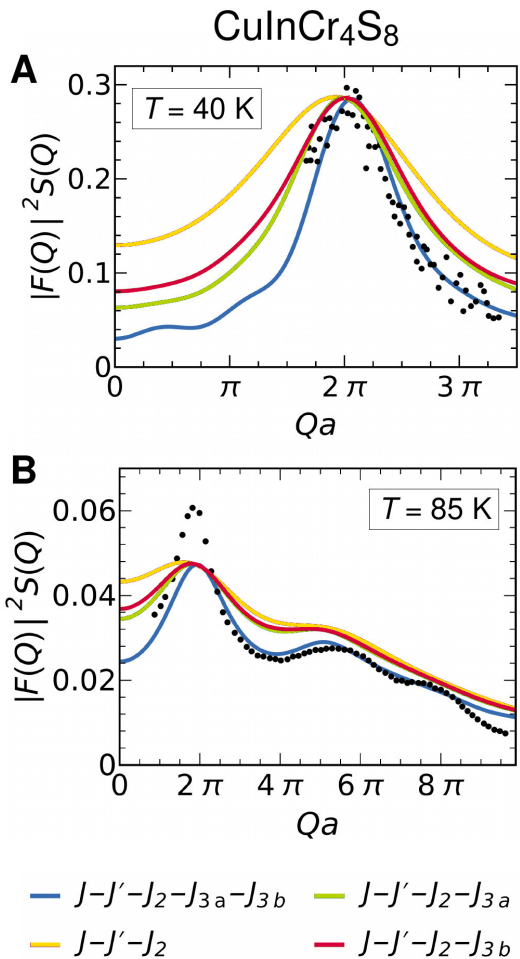

FIG. S6. **Importance of subleading interactions.** The comparison of $|F(Q)|^2 S(Q)$ for different combinations of $J$, $J'$, $J_2$, $J_{3a}$, $J_{3b}$ with the experimental data obtained by Plumier *et al.* (39) at (**A**) 40K and (**B**) 85K for $CuInCr_4S_8$. The experimental data is shown in black dots. From both the plots, it is very evident that the best match with the experimental data is *only* obtained only with the full $J$-$J'$-$J_2$-$J_{3a}$-$J_{3b}$ model. This further validates the value of exchange couplings used in our calculations regarding $CuInCr_4S_8$.

## S3. LUTTINGER-TISZA CALCULATIONS

We employ the Luttinger-Tisza method (53,54,61) to better understand the classical ground states of the breathing pyrochlore Hamiltonians. Figure S7 shows the evolution of the eigenvalues of the Luttinger-Tisza matrix for various breathing anisotropies $J'/J$. The minimum of the resulting bands identifies the ordering vector of the classical Hamiltonian in reciprocal space. The minima can also be degenerate, and the method allows us to identify one-, two- or even three-dimensional classical degeneracies of the ground state manifold.

Relevant for the chromium oxide spinels, figs. S7A to D show the eigenvalues for antiferromagnetic $J$ and $J'$. The lowest Luttinger-Tisza band is completely flat, indicating a three-dimensional manifold of degenerate ground states. Decreasing $J$ as compared to $J'$ in figs. S7B, C and D only shrinks the band width but does not lift the degeneracy of the lowest band. This explains why the typical bow-tie pattern in the magnetic susceptibility is nearly independent on the $J'/J$ ratio.

For a better understanding of the chromium sulfide spinels, figs. S7E to H show the case of a ferromagnetic large tetrahedron $J'$ together with several antiferromagnetic $J$ values. In this case, the minima of the lowest Luttinger-Tisza band correspond to the lines marked in red, with the ground state displaying a one-dimensional sub-extensive degeneracy. Again, this degeneracy is independent of the value of the $J'/J$ ratio. This explains why the three sulfides show very similar susceptibilities even though their Hamiltonians are rather distinct.

## S4. CLASSICAL MONTE CARLO RESULTS

Figure S8 shows the static spin structure factor as obtained from classical Monte Carlo simulations for $CuInCr_4Se_8$ at six different temperatures. In the vicinity of the ordering temperature $T/\bar{J} = 0.20$ the spectral weight is distributed in a nonuniform fashion being concentrated around the Bragg peaks of the incipient incommensurate spiral order. With increasing temperatures, the overall intensity of $S(\mathbf{q})$ decreases but the distribution of spectral weight becomes progressively homogeneous over a ring-like shape in the $[hk0]$ and $[hhl]$ planes. At temperatures of $\sim 1.5 T_{\text{ord}}$ we observe (i) a near uniform distribution of $S(\mathbf{q})$ along a ring-like shape, (ii) a concentrated distribution of $S(\mathbf{q})$ in a ring-like pattern of small width, and (iii) well-separated spiral rings. The existence of these features points to the appearance of a temperature regime hosting an approximate spiral spin liquid.

## S5. BASIS EFFECTS IN THE STRUCTURE FACTOR

We investigate the origin of the calculated susceptibilities of the chromium sulfides shown in Figs. 3G-I of the main text and figs. S4D-I. For this purpose, we use the Luttinger-Tisza method (53,54,61) to calculate the structure factor of the antiferromagnetic Heisenberg Hamiltonian on the fcc lattice. The result is shown in fig. S5C. Now, we have to take into account the fact that the classical $S = 6$ magnetic moment on the fcc lattice is not point-like but instead has an internal structure – the moment being composed of six $S = 3/2$ moments arranged on a tetrahedron constituting a basis. Thus, the complete structure factor of the pyrochlore lattice is now determined by multiplying the spin structure factor on the fcc lattice with the square of the absolute value of the

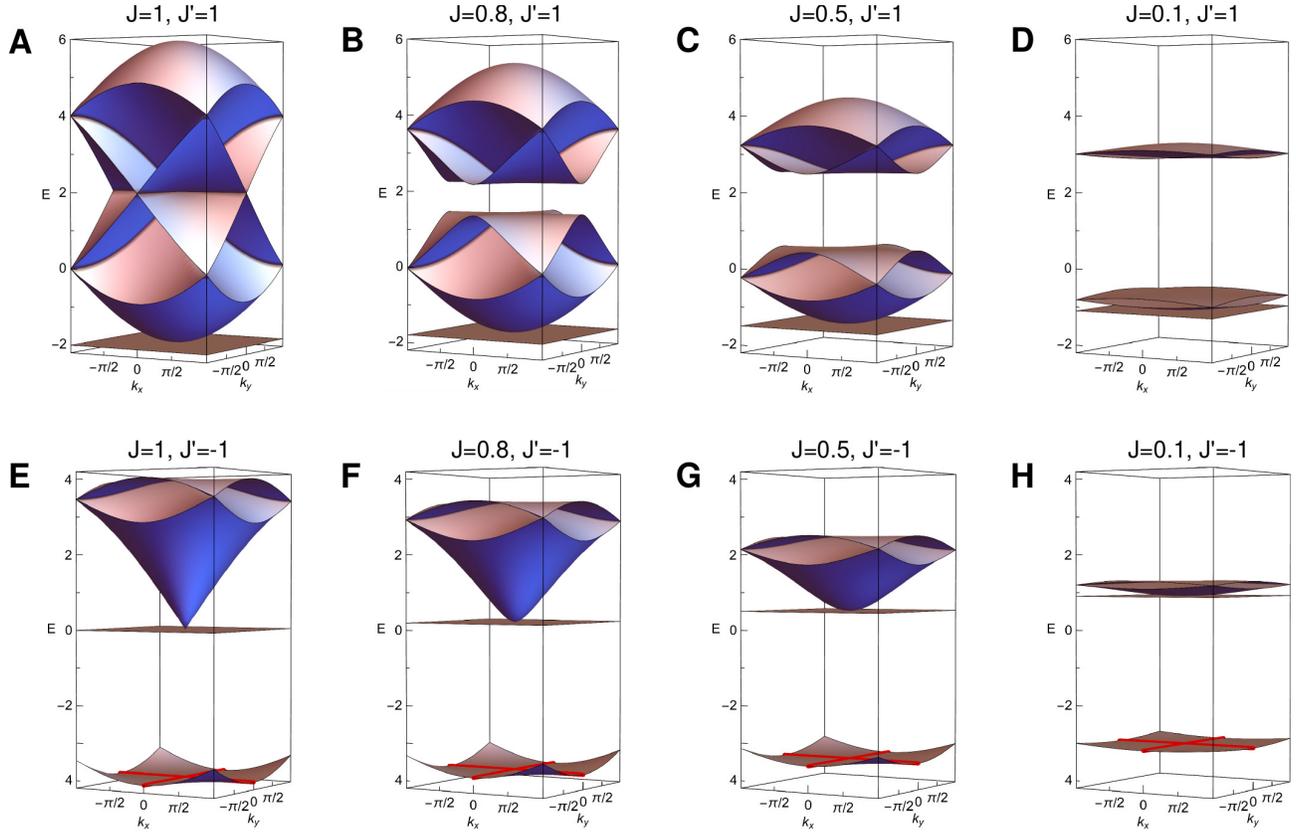

FIG. S7. **Luttinger-Tisza results.** Eigenvalues of the Luttinger-Tisza matrix for the breathing pyrochlore Hamiltonian with nearest-neighbor exchanges $J$ and $J'$. (**A**)-(**D**) Antiferromagnetic case for constant $J' = 1$ and decreasing $J$. (**E**)-(**H**) Ferromagnetic $J' = -1$ with decreasing antiferromagnetic $J$. All plots show the $(h, k, 0)$ plane, with $k_z$ chosen to be zero.

Fourier transform of the magnetic basis

$$f(\mathbf{q}) = \frac{1}{4}\left(1 + e^{-i\frac{a}{4}(q_x+q_y)} + e^{-i\frac{a}{4}(q_x+q_z)} + e^{-i\frac{a}{4}(q_y+q_z)}\right). \quad (S1)$$

The susceptibility profiles therefore acquire a factor of $|f(\mathbf{q})|^2$ between fcc and pyrochlore lattice:

$$S_{\mathrm{Pyro}}(\mathbf{q}) = |f(\mathbf{q})|^2 S_{\mathrm{fcc}}(\mathbf{q}). \quad (S2)$$

The classical spin susceptibility of a fcc nearest-neighbor Heisenberg antiferromagnet, relevant to the sulfides consists of degenerate maxima in the $[hk0]$ planes along the lines $\mathbf{q} = \frac{2\pi}{a}(\pm n, \delta, 0)$ and $\mathbf{q} = \frac{2\pi}{a}(\delta, \pm n, 0)$, which form edge sharing squares with side length $4\pi/a$ in reciprocal space, as shown in fig. S5C. The factor $|f(\mathbf{q})|^2$, shown in fig. S5D is 1 at $(0, 0, 0)$, and zero on the lines $\mathbf{q} = \frac{4\pi}{a}(\pm n, \delta, 0)$ and $\mathbf{q} = \frac{4\pi}{a}(\delta, \pm n, 0)$, and thus $8\pi$ periodic. This implies that every second row and column of squares in the fcc susceptibility is suppressed by the sublattice structure (see fig. S5E), this giving rise to a pattern of disconnected squares of maxima on the pyrochlore lattice which are also modulated in their intensity. Indeed, this is precisely what PFFRG finds for the chromium sulfide breathing pyrochlores (see Figs. 3G-I of the main text and figs. S4D-I).

7

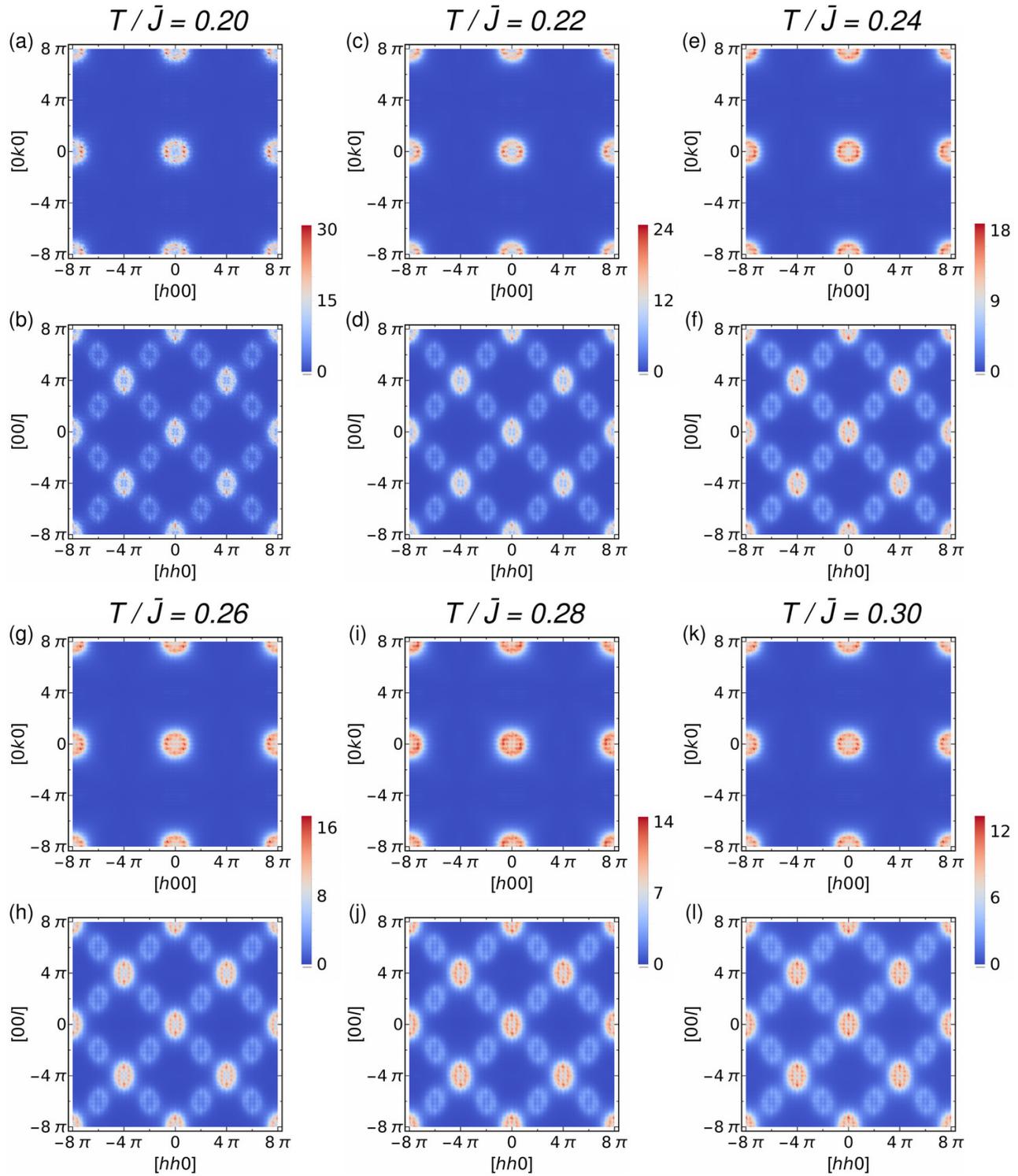

FIG. S8. **Classical Monte Carlo results.** The spin (not neutron) structure factor for $CuInCr_4Se_8$ obtained using classical Monte Carlo simulations (system of $32 \times 32 \times 32 \times 16 = 524288$ spins) projected on $[hk0]$ (first row) and $[hhl]$ (second row) planes. Susceptibility units are $1/\bar{J}$ ($\bar{J} = \sqrt{J^2 + J'^2}$). At the ordering temperature $T/\bar{J} = 0.20$, the Bragg peaks at $\mathbf{q} = (0.40 \pm 0.04)$ are clearly visible, however, at progressively higher temperatures the spectral weight gradually becomes more uniformly distributed leading to the appearance of an approximate spiral spin liquid. The axes are in units of $1/a$, $a = 10.5735$ Å (7).



\end{document}